%% file: smc.tex
\documentclass[11pt]{article}
\usepackage{aaspp4}
\usepackage{rotating}
\usepackage{epsfig}

\newcommand{\lmax}{$\lambda_{max}\ $}
\newcommand{\llmax}{$\lambda_{max}$}
\newcommand{\mic}{$\micron$}
\newcommand{\mmic}{$\micron\ $}
\newcommand{\clau}{$\micron^{-1}$}
\newcommand{\ccla}{$\micron^{-1}\ $}
\newcommand{\lamb}{$\lambda$}

\newcommand{\pmax}{$P_{max}\ $}
\newcommand{\qui}{$\chi^2$\ }
\newcommand{\qqui}{$\chi^2$}
\newcommand{\ppmax}{$P_{max}$}
\def\gtsim{\hbox{\raise.5ex\hbox{$>$}\llap{\lower.5ex\hbox{$\sim$}}}} 
\def\ltsim{\hbox{\raise.5ex\hbox{$<$}\llap{\lower.5ex\hbox{$\sim$}}}}

\tighten
\received{}
\accepted{}
\lefthead{Rodrigues et al.}
\righthead{Interstellar Dust in the SMC}

\slugcomment{Accepted by the Astrophysical Journal}

\begin{document}

\title{Dust in the Small Magellanic Cloud:\\
     Interstellar Polarization and Extinction}

\author{C. V. Rodrigues\altaffilmark{1} and A. M. 
Magalh\~aes\altaffilmark{2,3,4,5}}
\affil{Instituto Astron\^omico e Geof\'\i sico - USP\\
     Caixa Postal 9638, S\~ao Paulo - SP 01065-970, Brazil\\
     \it{e-mail: claudia@das.inpe.br, mario@argus.iagusp.usp.br}}

\author{G. V. Coyne, S.~J.}
\affil{Vatican Observatory\\
     V-00120 Vatican City State, Rome, Italy\\
     \it{e-mail: gcoyne@as.arizona.edu}}

\and
\author{V. Piirola}
\affil{Tuorla Observatory - University of Turku\\
	SF-21500 Piikki\"o, Finland\\
	\it{e-mail: piirola@sara.cc.utu.fi}}

% Notice that each of these authors has alternate affiliations, which
% are identified by the \altaffilmark after each name.  The actual alternate
% affiliation information is typeset in footnotes at the bottom of the
% first page, and the text itself is specified in \altaffiltext commands.
% There is a separate \altaffiltext for each alternate affiliation
% indicated above.

\altaffiltext{1}{Now at Inst. Nacional de Pesquisas Espaciais-INPE,
     Divis\~ao de Astrof\'\i sica-DAS, Caixa Postal 515,
     S\~ao Jos\'e dos Campos - SP 12201-970, Brazil}
\altaffiltext{2}{Guest observer at IUE}
\altaffiltext{3}{Visiting Astronomer, Cerro Tololo Inter-American Observatory.
CTIO is operated by AURA, Inc. under contract to the National Science
Foundation}
\altaffiltext{4}{Visiting Astronomer, Complejo Astronomico El Leoncito, San Juan,
Argentina}
\altaffiltext{5}{Visiting Astronomer, European Southern Observatory, La Silla,
Chile}

% The abstract environment prints out the receipt and acceptance dates
% if they are relevant for the journal style.  For the aasms style, they
% will print out as horizontal rules for the editorial staff to type
% on, so long as the author does not include \received and \accepted
% commands.  This should not be done, since \received and \accepted dates
% are not known to the author.

\begin{abstract}

The typical extinction curve for the Small Magellanic Cloud 
(SMC), in contrast to that for the Galaxy, has no bump at 2175 \AA\
and has a steeper rise into the far ultraviolet. For the Galaxy
the interpretation of the extinction and, therefore, the dust
content of the interstellar medium has been greatly assisted by
measurements of the wavelength dependence of the polarization. To
the present no such measurements existed for the SMC.

Therefore,
to further elucidate the dust properties in the SMC we have for
the first time measured linear polarization in five colors in the
optical region of the spectrum for a sample of reddened stars.
For two of these stars, for which there were no existing UV
spectrophotometric measurements, but for which we measured a
relatively large polarization, we have also obtained 
data from the International Ultraviolet Explorer (IUE)
in order to study the extinction.
With the help of parametrization, we attempted to correlate
the SMC extinction and polarization data. In addition, we
performed dust model fits to both extinction and polarization
using silicate and graphite or amorphous carbon spheres and silicate cylinders.
The size distribution for the cylinders is taken from a fit to
the polarization and we introduce the notion of volume continuity between
this and the silicate sphere size distribution.

The main results are: (1) the wavelength of maximum polarization, 
\llmax, in the SMC is typically smaller than that in the 
Galaxy; (2) however, AZV 456, which shows the UV extinction bump,
has a \lmax typical of that in the Galaxy, its polarization curve
is narrower, its bump is shifted to shorter wavelengths
as compared to the Galaxy and its UV extinction does not conform to
the Galactic analytical interpolation curve based on the ratio of total
to selective extinction;
(3) the 'typical', monotonic SMC extinction curve
can be best fit with amorphous carbon and silicate grains;
(4) the extinction towards AZV456 may only be explained by assuming a larger
gas-to-dust ratio than the observed N(HI)/A(V) value, with a small amount of
the available carbon in graphite form;
(5) from an analysis of both the
extinction and polarization data and our model fits it appears that the SMC has
typically smaller grains than those in the Galaxy.

The absence of the extinction bump in the SMC has generally been thought to
imply a lower amount or even an absence of carbon in solid form in the
SMC compared to the Galaxy.
Our results show that the size distribution, and not only the carbon
abundance, is
different in the SMC as compared to the Galaxy. In addition, and contrary to
previous findings, another component
besides silicates is indeed needed to provide a sizeable part of the
observed extinction towards the SMC.
Using the SMC as a laboratory for studying the solid component of the
interstellar medium, we also discuss some of the implications of our
results in view of proposed interstellar dust models.

\end{abstract}


% That's it for the front matter.  On to the main body of the paper.
% We'll only put in tutorial remarks at the beginning of each section
% so you can see entire sections together.
%
% In the first two sections, you should notice the use of the LaTeX \cite
% command to identify citations.  The citations are tied to the
% reference list via symbolic tags.  We have chosen the first three
% characters of the first author's name plus the last two numeral of the
% year of publication.  The corresponding reference has a \bibitem
% command in the reference list below.
%
% Please go to the LaTeX manual for a complete description of the
% \cite-\bibitem mechanism.

\section{Introduction}

     The interstellar medium (ISM) in galaxies is indicative of 
their evolutionary state and of their stellar populations. For 
instance, the Galaxy contains four times more heavy elements than 
the Large Magellanic Cloud (LMC) and ten times more than the 
Small Magellanic Cloud (SMC) (Wheeler, Sneden \& Truran 1989).
The dust content of the ISM in the LMC (Koorneef 1982;
Fitzpatrick 1985a) and SMC (Bouchet et al 1985; Fitzpatrick
1985b) is less than in the Galaxy and the dust is qualitatively
different. In particular, the SMC has a steeper far ultraviolet
(FUV) extinction curve (Pr\'evot et al. 1984); it typically has
no ultraviolet (UV) extinction bump. The SMC has also a weak
infrared (IR) emission at 12 \mmic and an intensity ratio
60\mic/100\mmic larger than that in the Galaxy (Lequeux 1989).
This would appear to indicate that star formation and evolution
have proceeded at a faster rate in the Galaxy than in the LMC or
SMC and/or that dust grains have formed there differently in a
different environment. In fact, there is some indication that the
efficiency of grain formation in the LMC and SMC may be slightly
lower, albeit similar, compared to that in the Galaxy (Clayton
\& Martin 1985). Also, from data on M31, it is suggested
that the abundance of small grains may be related to star
formation rates (Xu \& Helou 1994).

     For these reasons and because of its proximity 
the SMC is an important environment in which to study dust
grains. Dust may cause several observable effects: it attenuates,
reddens, scatters and polarizes starlight; it absorbs and reemits
radiation; it may bring about a depletion of the gas content of
the ISM and may serve as a catalyst for the formation of
molecules in the ISM. In the case of the SMC, the higher FUV
radiation field together with the lower dust content produces an
increased photodissociation of molecules that affects the
properties of molecular clouds (Lequeux et al 1994). The SMC is
thus an excellent laboratory to study dust in an environment
quite distinct from that of the Galaxy, while at the same time
allowing us to test dust models which have been proposed. We
concentrate our study on the reddening and the polarization due
to dust in the SMC.

     It has been known for some time from IUE observations that
the UV and FUV extinction relative to the visual and the IR is
much larger in both the LMC (Nandy et al 1981; Koorneef \& Code
1981; Fitzpatrick 1985a, 1986) and in the SMC (Pr\'evot et al
1984) than it is in the Galaxy and the effect is larger for the
SMC than it is for the LMC. The SMC is also noteworthy for the
lack of the extinction bump at 2175 A (see review by Fitzpatrick
1989). Interstellar polarization in connection with extinction in
the LMC context has been studied by Clayton, Martin \& Thomson
(1983), Clayton \& Martin (1985) and Clayton et al. (1996).

     Interestingly, the SMC star AZV 456 (AZV = Azzopardi \&
Vigneau 1982) shows an extinction and a gas-to-dust ratio
(Lequeux et al. 1984) similar to  the average values in the
Galaxy. The interstellar lines in this direction have typical
velocities for the SMC, so that the extinction is probably not 
foreground (Lequeux et al. 1984; Martin, Maurice \& Lequeux
1989). There is an IR extended source in the SMC located at
the same coordinates as AZV 456 (LI-SMC 190: Schwering \&
Israel 1989), which also indicates that the extinction is within
the SMC. Lequeux (1994) has raised the possibility that
AZV~456 may have, in fact, a higher gas-to-dust ratio due
to a possibly undetected amount of H$_{2}$. In section \ref{av456}
we show that such higher ratio is indeed required if the extinction
towards this object is appropriately fit.

     Classical models for fitting the wavelength dependence of
the extinction by dust grains in the ISM of galaxies generally
take into account the chemical composition of the grains and
their size distribution. Bromage \& Nandy (1983) and Pei (1992)
studied the SMC 'typical' extinction curve using the model of
Mathis, Rumpl \& Nordsiek 1977 (henceforth, MRN) with only
spherical particles. They concluded that, by simply lowering the
quantity of graphite grains relative to silicates in the Galactic
models, it was possible to fit the {\it wavelength dependence} of the
SMC extinction without changes
in the sizes used to obtain the Galactic curve. Pei (1992) and
Maccioni \& Perinotto (1994) have studied the extinction in the
LMC and the latter noted from their fits to the extinction that
no unique solution could be obtained for the grain sizes and
abundance ratios.

     In the Galaxy it is known that the wavelength at which the
polarization is maximum is related to the grain size (Coyne,
Gehrels \& Serkowski 1974). It would, therefore, be very useful
to have this information for the SMC. No wavelength dependent
polarization measurements have been yet published on the SMC. In
this paper we present the first such observations and analyze
them in light of the extinction, for which we also report a few
new IUE observations. In addition, we present some model fits to
these observations.

\section{Data}

\subsection{Polarization Data}
\label{polarization_data}

\subsubsection{Observations}

Polarimetric observations have been obtained during several 
observing runs at various observatories and with various
polarimeters. The observing runs were: 1983 with the MINIPOL 
polarimeter (Frecker \& Serkowski 1976) on the 1.5 meter 
telescope at the Cerro Tololo Interamerican Observatory (CTIO); 
1987 with the PISCO polarimeter (Stahl et al. 1986) on the 2.2. 
meter telescope of the European Southern Observatory (ESO); and 
1986, October 1988, November 1988 and 1989 with the VATPOL 
polarimeter (Magalh\~aes, Benedetti \& Roland 1984) on the 2.15
meter telescope  at the Complejo Astronomico El Leoncito
(CASLEO). About seventy percent of the observations were made at
CASLEO.

     VATPOL and MINIPOL provide on-line data reduction after each 
integration and/or after a series of integrations (the
integration time can be freely selected). This data reduction 
consists of a least squares fit to a double cosine curve of the 
counts from two photomultiplier tubes obtained as a function
of the position of a rotating half-wave plate. The mean error of
the polarization is calculated from the actual deviations of the
two counts from the double cosine curve; this error is typically
consistent with photon-noise error, as detailed by Magalh\~aes et
al. (1984). The data from PISCO were obtained in FITS format and
we wrote a special microcomputer program to calculate the
polarization from the star and sky counts in a way similar to
that of the other two instruments.

     All data have been corrected for instrumental effects and 
have been standardized. Unpolarized standard stars were measured 
to obtain corrections for instrumental polarization. Nightly
measurements were made on highly polarized standard stars and
also with a Glan prism in the beam in order to obtain the
polarizing efficiencies and to standardize the polarization
position angles to the equatorial system. The polarizing
efficiencies were typically 98 to 99 \% for VATPOL and MINIPOL.
For the measurements with PISCO the efficiencies were provided to
us by Hugo Schwarz of ESO. Repeated runs at CASLEO showed all
corrections measured there to be stable. Corrections for bias in
the linear polarization, which depend on the ratio of the error
to the percentage polarization (Clarke \& Stewart 1986), were
also applied. We note, however, that all calculations were made
using the Stokes parameters.

     We combine the results from the three instrumental systems 
in the following way. For each star we calculate first the
weighted average of the measurements made with each filter in 
each of the instrumental systems. We then determine the weighted
mean of those averages. The results, both observed and corrected
for foreground polarization (see sec. 2.1.2) are given in Table \ref{tab_pol}
where the respective columns are: (1) star number from Azzopardi
\& Vigneau (1982, AZV82); (2) star number from Sanduleak (1968,
1969); (3)-(5): the observed polarization, equal to the weighted
mean polarization over the averages obtained from each observing
run, the standard deviation of the mean and the polarization
position angle in the equatorial system, respectively; (6)-(8):
same as columns (3)-(5) but for the foreground corrected
polarization; (9) the effective wavelength (averaged over
the values for each observing run, by weighting with the mean
polarization errors). The foreground-corrected polarization
measurements are plotted in Fig. \ref{fig_serk}.

     The targets reported in Table \ref{tab_pol} have been selected for
multifilter polarimetry from our on-going program to map the
magnetic field structure of the SMC (Magalh\~aes et al. 1990).
The sample was built from the AZV82 catalog, avoiding stars with
emission line spectra. This survey is presently being conducted
with CCD imaging polarimetry (Magalh\~aes et al. 1996) and the
results will be reported elsewhere.

\subsubsection{Foreground Polarization Corrections}
\label{for_pol}

     It is necessary to correct the measured polarizations by 
subtracting the polarization foreground to the SMC due to 
dust in the Galaxy. It is expected that these corrections are 
important, since the interstellar reddening intrinsic to the SMC 
is small. We have selected published, unfiltered polarization
data (Mathewson \& Ford 1970; Schmidt 1976) on about 40 stars
farther than 400 pc from the Sun in the direction of the stars
measured in the SMC. McNamara \& Feltz (1980) have shown that
most of the foreground extinction to the SMC occurs within 400 pc
from the Sun. Also by this selection we include most of the dust
in the Galactic plane for which the scale height is 120 pc
(Burton et al. 1986).

     As a check on this procedure, we have also estimated the
foreground polarization in two additional ways. The first of
these was to average the polarization data of our survey sample
(Magalh\~aes et al. 1990, Rodrigues 1992) for stars in the SMC
which showed an observed polarization equal to or less than
0.4\%. We have also averaged, again using data from our survey,
polarization data for SMC stars which had an estimated color
excess less than 0.09 mag. This upper limit for the foreground
reddening towards the SMC is suggested by Schwering (1988);
McNamara \& Feltz (1980) and Bessel (1991) have suggested lower
values (0.02 mag and 0.04 mag-0.06 mag, respectively). We have
estimated the color excess for the SMC stars in our sample using
the intrinsic colors from Fitzgerald (1970) and Brunet (1975).
Both of these methods gave results that were entirely consistent
with the estimates from the Galactic foreground objects (Table \ref{tab_cor})
described in the previous paragraph and which we used for
correcting the SMC data.

     Table \ref{tab_cor} gives the adopted corrections for the foreground
polarization towards the various regions in the SMC defined
by Schmidt (1976). Our program stars are located as follows:
region I, AZV 20, 126, 221; region II, AZV 211, 215, 398; region
III, AZV 456. The estimated values for the foreground
polarization were taken as valid for the V filter. We used the  
Serkowski law, with \llmax=0.55\mic, to estimate the contribution
in the other filters. The errors of the corrected polarization
values include the increase in uncertainty arising from this  
correction. These are the values in Table \ref{tab_pol}. We tested
the influence of changes in \lmax on the above estimates and on
the resulting corrected data and found them to be insignificant.

\subsection{UV spectroscopic data}

     Our UV sample consisted of three reddened stars, AZV 20, AZV
126 and AZV 211, and comparison stars necessary to obtain the 
extinction curves (see sec. 2.2.2). AZV 211 was our primary
target because it had shown a well determined polarization curve
with a small \lmax (Fig. \ref{fig_serk} and Table \ref{tab_serk}). We have included in our
reduction the data already published on AZV 398 (Pr\'evot et al. 
1984) and on AZV 456 (Lequeux et al. 1984) in order to test our 
procedure and to give homogeneity to the sample studied in the 
following sections. The relevant data on the reddened stars and
on the comparison stars are given in Tables \ref{tab_prog} and \ref{tab_comp}
respectively.

     The UV spectral data were obtained with the International 
Ultraviolet Explorer (IUE, Boggess 1978a,b; Kondo 1987) in two 
runs: November, 1990 and September, 1991. The images were 
obtained with three cameras: the Small Wavelength Prime
(SWP, 1200 to 2000 \AA); the Large Wavelength Prime (LWP, 2000 to
3200 \AA) and the Large Wavelength Redundant (LWR, 2000 to 3200
\AA). We have also used some images from the IUE data archive
(see last column of Tables  \ref{tab_prog} and \ref{tab_comp}).
The images are
unidimensional vectors with a sampling of 1 \AA.

\subsubsection{Obtaining the combined spectrum of each star}
\label{comp_spec}

     Each star has been observed, sometimes more than once, in 
each of the two wavelength ranges. The reductions were made using
the RDAF and IUEIDL packages at the University of
Wisconsin-Madison. The first procedure in the  data reduction is
to combine all the images of a star to one spectrum. Corrections
for the time degradation of the camera sensitivity (Bohlin \&
Grillmair 1988a,b) are made first. At the time this reduction was
done only corrections up to 1988 were available, so to correct
the data obtained after that time we had to use an extrapolation.
Also the cameras do not have the same efficiencies, so there may
be a discontinuity in the overlap region between two spectra. It
is generally assumed that the two spectra must have the same flux
in the overlap region. As we are interested only in the ratio
between spectra of the program and the comparison stars (see eq.
\ref{eq_ext}) we have not corrected for this effect. The combined spectra
did not in fact present any discontinuities.

\subsubsection{Determination of the extinction curves}
\label{ext_det}

     We have obtained the extinction curves using the pair-method 
(Fitzpatrick \& Massa 1986) which consists in comparing two stars 
of the same spectral type, but with different reddening. The
assumption is that the program and comparison stars have exactly 
the same intrinsic spectra, the observed difference being due to 
the foreground interstellar medium. We estimated the error in 
this assumption by using different comparison stars.

     From the B and V magnitudes and the fluxes, $\phi_i$, 
the normalized extinction is given by:

\begin{equation}
\label{eq_ext}
\frac{E(\lambda-V)}{E(B-V)} = \frac{2.5 
\log(\frac{\phi_c}{\phi_r})-V_r+V_c}
{(B-V)_r-(B-V)_c}
\end{equation}

\noindent where the subscripts $r$ and $c$ mean {\it reddened}
and
{\it comparison star}, 
respectively. Great care has been exercised in selecting
appropriate comparison stars, specially for AZV~211 due to its
relatively late spectral type. The comparison stars have been
chosen from among SMC stars only, using the AZV82 catalog and
excluding emission line stars, which are often variable and
present anomalous color excesses due to circumstellar material.
We also chose these unreddened stars within a spectral sub-class
from AZV~211 with the purpose of matching as best we could this
star's spectral type. Further, we chose to stay with comparison
objects within about a magnitude of AZV~211. By using SMC stars,
we also minimize the effect of metallicity differences between
comparison stars and reddened stars, since two stars with the
same optical spectral type but different metallicities may have
different UV spectra. In addition, by using unreddened SMC
objects the foreground Galactic extinction is canceled if it is
homogeneous across the SMC angular field. We have later examined
the UV lines to detect possible mismatches.

     AZV 20 and AZV 211 have the same spectral type (A0 Ia), so 
we have used the same comparison stars, observed by us, for both 
of them (see Tables  \ref{tab_prog} and \ref{tab_comp}).
Stars of spectral type A0 Ia 
present some difficulties. The UV spectra change rapidly with
spectral type so that a mismatch has a much greater effect than
it does for stars of earlier type. Furthermore, since stars of
this spectral type are cooler and their UV flux lower, it is more
difficult to get a good signal-noise ratio. AZV 126 has a
spectral type (B0Iw) very similar to that of AZV 398 (B2) and AZV
456 (B0-1), so there is an overlap in the comparison  stars,
taken from the IUE data archive, for all three (see Table \ref{tab_comp}).
AZV 126 however is the least reddened of these three stars; the
comparison stars were selected in order to provide the largest
possible difference between program and comparison stars (see
next paragraph). The final list of comparison stars for each
program star is: for AZV 20 and 211, the comparison stars were
AZV 161, 270, 504 and SK 194; for AZV 126, the comparison stars
were AZV 61, 317 and 454; for AZV 398 and AZV 456, the comparison
stars were AZV 235, 242, 289, 317 and 488.

     As noted above, the IUE spectra have 1 \AA~ sampling. The
extinction calculated with such a small wavelength step usually
has a very poor signal-to-noise ratio. So we have also formed 
combined spectra with bins of 80 \AA. The flux value in each 80
\AA~ bin is assumed to be the sum of the fluxes in the 1 \AA~
bins. The extinction curves are shown in Figs. \ref{fig_ext} and \ref{fig_fm}.
The
curves with 1 \AA~ bins (Fig. \ref{fig_ext}) are useful to identify spectral
type mismatches and the noisier regions. The curves in the
figures are the weighted average of the curves using different
comparison stars and the error bars in the 80 \AA~ curves (Fig.
\ref{fig_fm}) are the average standard deviation. The weights used were the
values of $\Delta(B-V)$, the difference between the program star
and comparison star colors given in Tables \ref{tab_prog}
and \ref{tab_comp}. In this way
we give greater weight to the reductions with the least reddened
comparison stars. Also, this way of weighting is appropriate for
errors dominated by measurement inaccuracies.

     For AZV 211 we initially determined extinction curves from
the four comparison stars listed above. We found, however, that
the curve using the comparison star AZV 161 was very different
than the other three curves, and we suspect that this comparison
star may be reddened. We have, therefore, excluded the curve
using AZV 161 from the average curve for AZV 211 and for AZV 20.
The resultant average curve for AZV 211 seems to follow the SMC
standard with no bump and an enhanced FUV extinction (Figs. 
\ref{fig_ext}c and \ref{fig_fm}c).

     AZV 20 has the noisiest extinction curve (see Fig.  \ref{fig_ext}a) in
the sample despite the fact that it has a large color excess (see 
Table \ref{tab_prog}). The systematic increase in error with increasing
frequency is due to the systematic decrease in signal-to-noise
(see Fig. \ref{fig_ext}a). Where the extinction curve is less noisy, the
values of the extinction are close to those for the typical SMC
extinction curve (compare, for example, Fig. \ref{fig_ext}a to Fig.
\ref{fig_ext}c and
Fig. \ref{fig_ext}d at ${\lambda}^{-1} \approx$ 5.5\clau). Its
binned extinction curve shows bin-to-bin fluctuations larger than
expected from the error bars.

     The extinction curves which we have redetermined for AZV 398
(Figs. \ref{fig_ext}d and \ref{fig_fm}d) and AZV 456
(Figs. \ref{fig_ext}e and \ref{fig_fm}e)
are in perfect agreement with those of Pr\'evot 
et al. (1984) and Lequeux et al. (1984), respectively.

The extinction for AZV 126 (Fig. \ref{fig_ext}b) will be discussed in
section \ref{param} below. AZV 215, for which we have
obtained polarization data (Tables \ref{tab_pol} and \ref{tab_serk}),
has spectra available in the IUE data bank. It has however
a very small $(B-V)$ 
value, comparable to possible comparison stars, so we could not 
determine a reliable extinction curve for it.

\section{Qualitative study of the optical polarization}
\subsection{Fits of the Serkowski-law to the SMC Polarization Data}
\label{serk_fit}

Serkowski (1973; see also Coyne et al. 1974) has shown that the
Galactic interstellar polarization can be described by the following
expression, 

\begin{equation}
\label{eq_serk}
P(\lambda ) = P_{max} \exp \left[-K
\ln^2\left(\frac{\lambda_{max}}{\lambda}\right)\right]
\end{equation}

where, from the observed  data, the parameters \pmax , K
and \lmax may be obtained. 
 
\ppmax, the maximum polarization, depends on the column density
of dust as well as the magnetic field structure and alignment
efficiency along the line of sight. \lmax is the wavelength where
\pmax occurs and it is related to size of the dust particles
(Coyne et al. 1974; Chini \& Kr\"ugel 1983). K describes the
width of the polarization curve.

     In his original work Serkowski has taken K as a constant
with a value of 1.15. Codina-Landaberry \& Magalh\~aes (1976)
have shown that K varied for different lines of sight.
Furthermore, from model fits they showed that K could be
interpreted as being related to changes in the grain size along
the line of sight. Wilking et al. (1980, 1982), using an extended
wavelength range that included the IR, suggested a linear
relation between K and \llmax. Whittet et al (1992), with an even
larger sample, have provided the following relation: 

\begin{equation}
\label{eq_whittet}
K = (1.66\pm 0.09) \lambda_{max} + (0.01\pm 0.05)\ . 
\end{equation}

     We have performed fits of the Serkowski relation to our SMC
data in two ways: (1) allowing K to be a free parameter; and (2)
using the above relation between K and \llmax. Admittedly, the
first approach results in larger uncertainties for the derived
parameters, especially K, but we felt that a first comparison
between the K values from the SMC data and those from Galactic
data would be of interest. Furthermore, a comparison between the
two methods might help to judge the reliability of the derived
parameters.
 
\begin{figure}
\epsfig{file=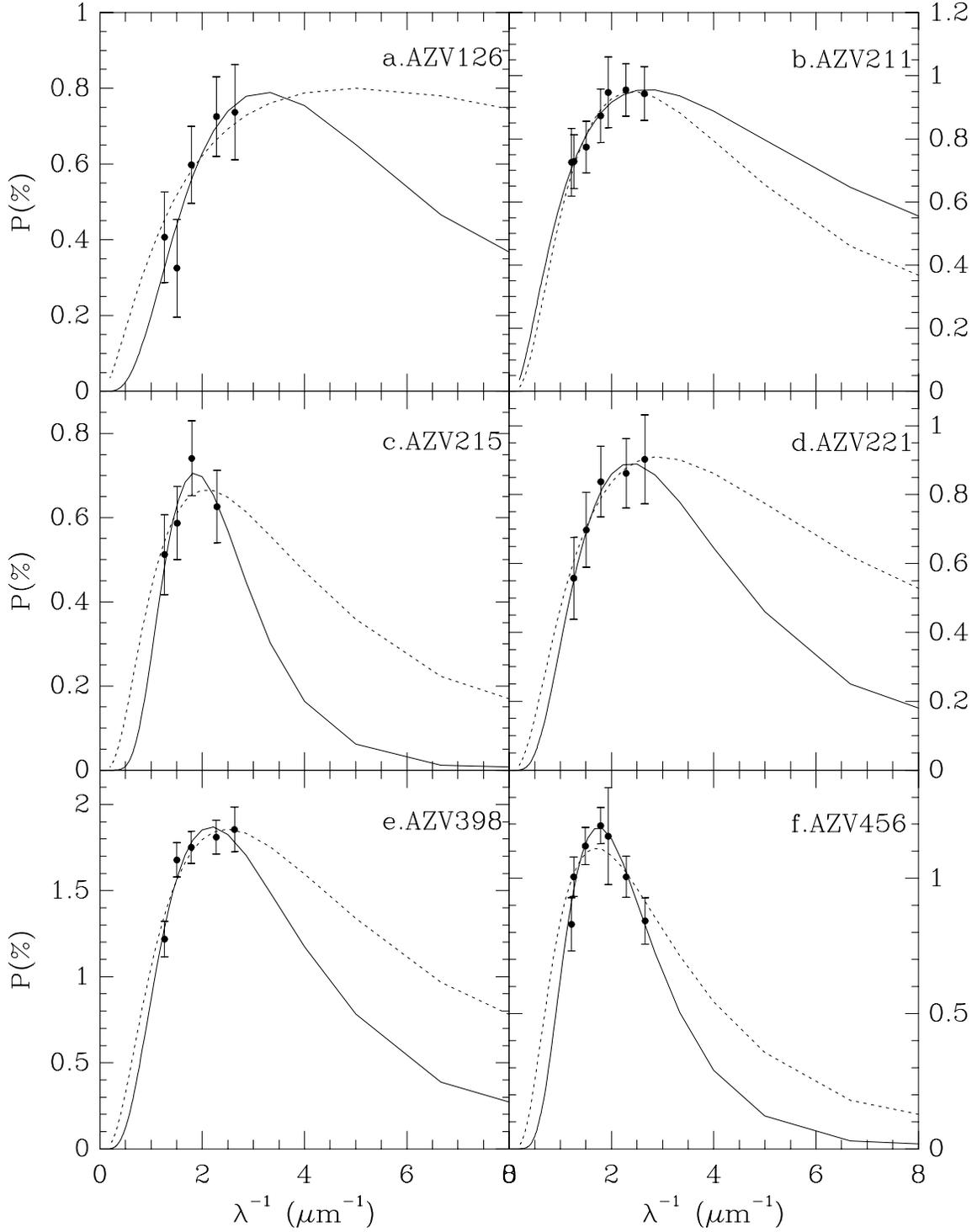,width=16cm}
\caption{Polarization for the SMC stars. Points with error bars
represent our foreground corrected data. The solid lines and
dashed lines show the 3- and 2-parameter Serkowski fits,
respectively.}
\label{fig_serk}
\end{figure}

\begin{figure}
\epsfig{file=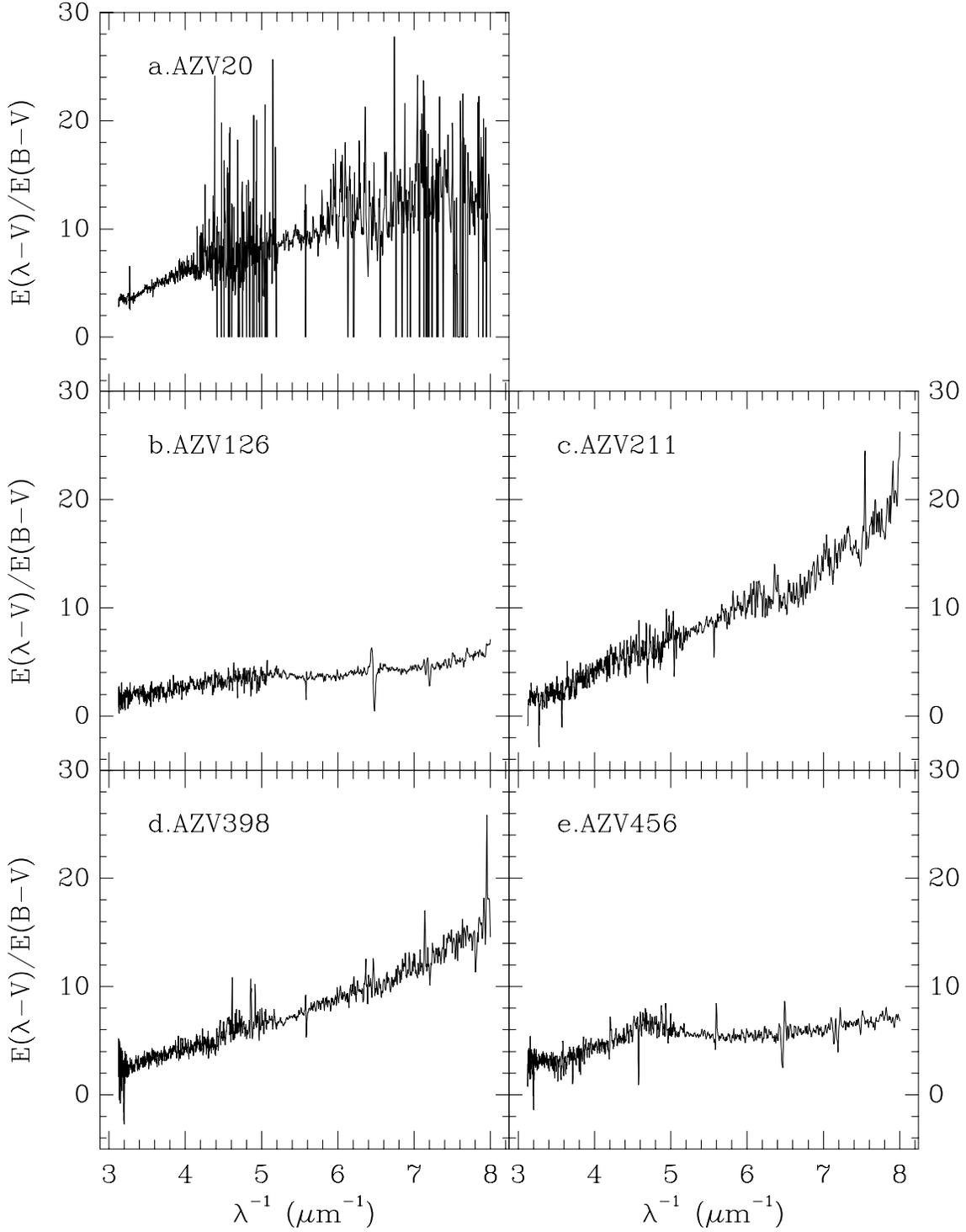,width=16cm}
\caption{Extinction curves of SMC stars obtained using the
pair-method and combined spectra with 1 \AA~ sampling (except for
5\AA\ sampling for AZV~126). The curves for AZV~398 and AZV~456
were derived using in part IUE images of these objects obtained
by Pr\'evot et al. (1984) and Lequeux et al. (1984).}
\label{fig_ext}
\end{figure}

\begin{figure}
\epsfig{file=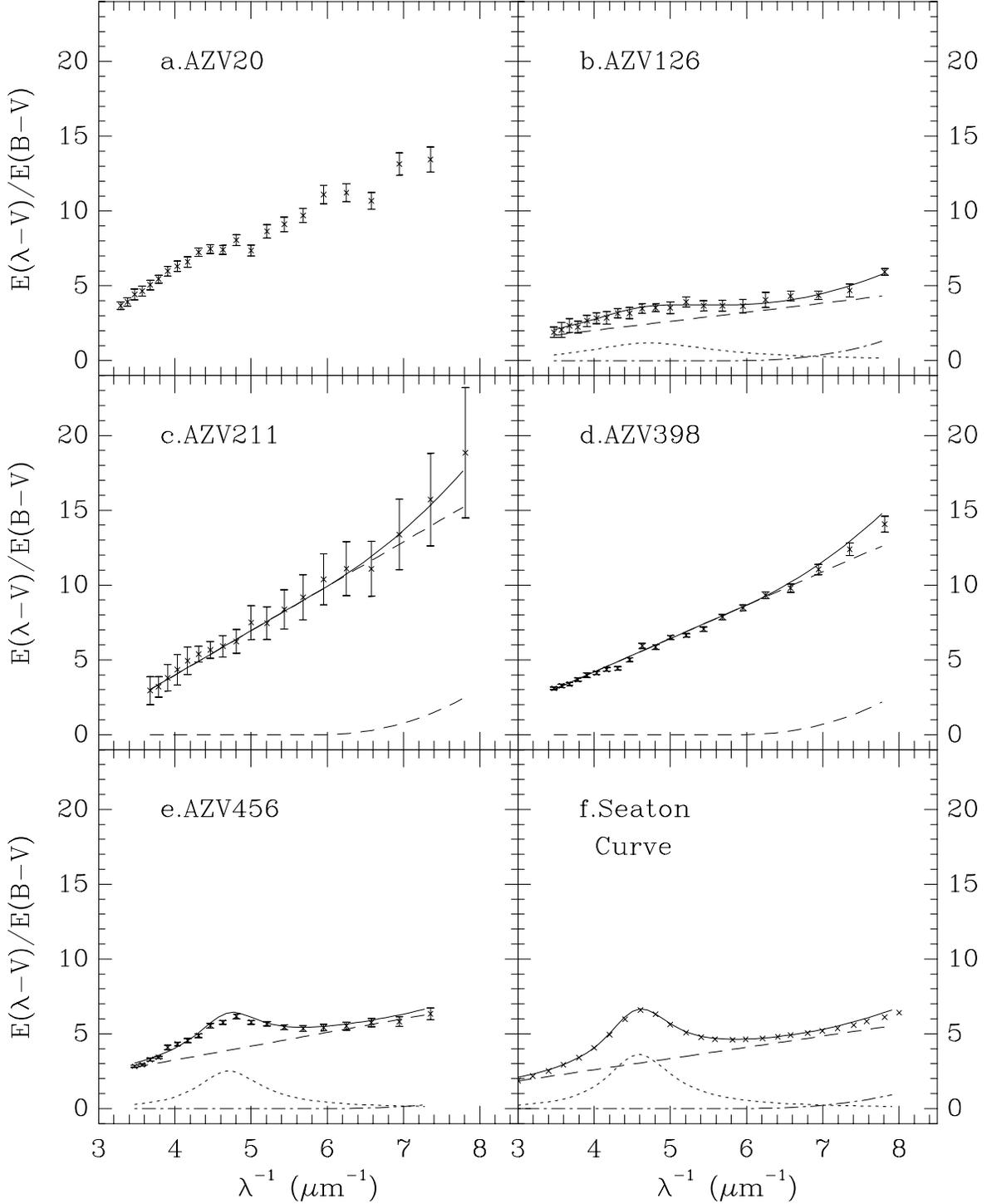,width=16cm}
\caption{Parametric fits, using the Fitzpatrick \& Massa (1986)
representation, to the extinction curves for: AZV 20, 126, 211,
398, 456 and the Seaton curve for the Galaxy. The fits (solid
line) were made to the extinction curve with a sampling of 1 \AA~
(except for 5~\AA\ sampling for AZV 126). The points with error
bars represent the extinction curve with a binning of 80 \AA.
The respective component curves are: dashed, the linear portion;
dotted, the Drude function; dashed-dotted, the polynomial
function for the UV and FUV. See eq. 5.}
\label{fig_fm}
\end{figure}

     Table \ref{tab_serk} gives the Serkowski fit parameters. In that table,
the last three columns show respectively the reduced $\chi^2$,
the associated probabilities and the degrees of freedom . The
actual fits are shown in Fig. \ref{fig_serk}. The SMC polarization data
can be well fit by the Serkowski relation by using either the 2-
or 3-parameter method. The only fit which is significantly
improved with three free parameters is that of AZV~456.
 
     The values of \lmax from the two methods agree well within
the errors. \pmax from both fits shows an even closer agreement.
Larger differences are found for the K  parameter, although they
may still be consistent within the large uncertainties.  AZV 215
has data points at only 4 wavelengths and hence the uncertainties
in K are particularly large.

     Table  \ref{tab_serk} shows that most of the SMC stars show \lmax smaller
than the Galactic average, 0.55\mic. In particular, this is true
for two of the stars with the best polarimetric signal-to-noise,
AZV~211 and AZV~398, which also have a typical SMC extinction
curve (sec. \ref{ext_det}). These results are
in sharp contrast, for instance, with those for the LMC (Clayton
\& Martin 1985). For their sample of stars with measured
extinction the smallest observed \lmax value is 0.52\mic, with a
median value of 0.58\mic. Our results will be further discussed
in sec. \ref{discussion}.

\begin{figure}
\epsfig{file=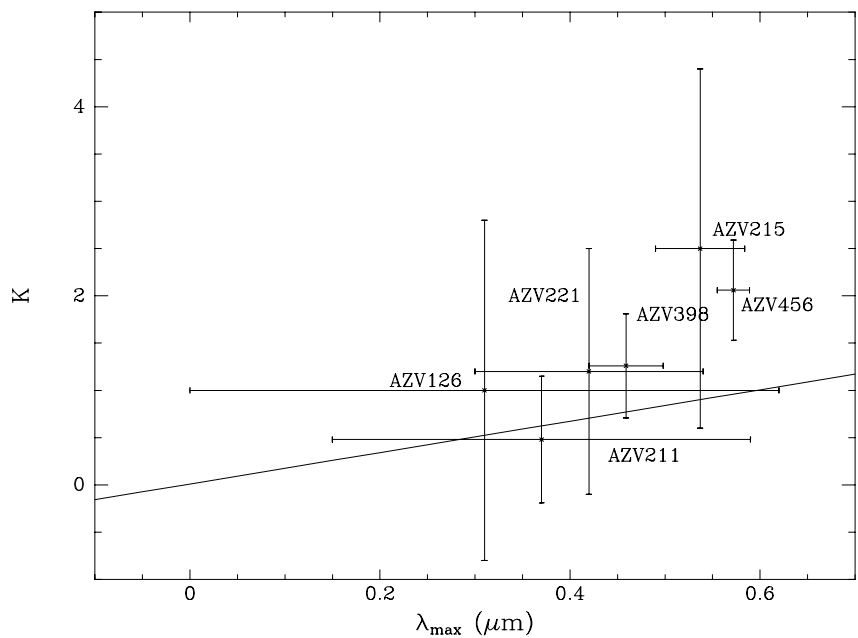,width=9.cm,angle=-90}
\caption{K versus \lmax for SMC stars (this work) from the
3-parameter fits. The solid line is the Galactic relationship
(Whittet et al. 1992).}
\label{k_lmax}
\end{figure}

AZV~456, which has a UV bump (sec. \ref{param}) and, to a certain degree,
AZV~215 show \lmax values close to the Galactic norm.
The 3-parameter fits (Table \ref{tab_serk} and
Fig. \ref{fig_serk}) indicate that their polarization curves are narrower than
those for the other stars. The fact that the Serkowski fit to the
'normal \llmax' polarization curve of AZV 456 is significantly
poorer when we use the Galactic relation between K and \lmax
may suggest that there is a different relationship
between these parameters in the SMC. This suggestion is somewhat
strengthened by a plot of the K vs. \lmax taken from the
three-parameter fits in Table \ref{tab_serk} and shown in
Fig. \ref{k_lmax}. While the K-\lmax
relation for the SMC is similar to the Galactic one, a steeper
slope is suggested. More data beyond the optical domain and for
more stars are needed to clarify this.

\subsection{Polarization and Visible Extinction}
\label{pol_ext}

     The maximum polarization towards a given line-of-sight is
related to the available amount of dust and depends on factors
such as the magnetic field direction, grain alignment efficiency  
and the polarizing efficiency of the grains. Empirically, it is
verified that for the Galaxy (Serkowski, Mathewson \& Ford
1975).

\begin{equation}
\label{eq_pmax}
P_{max} \leq 9.0 E(B-V)\ .
\end{equation}

\begin{figure}
\epsfig{file=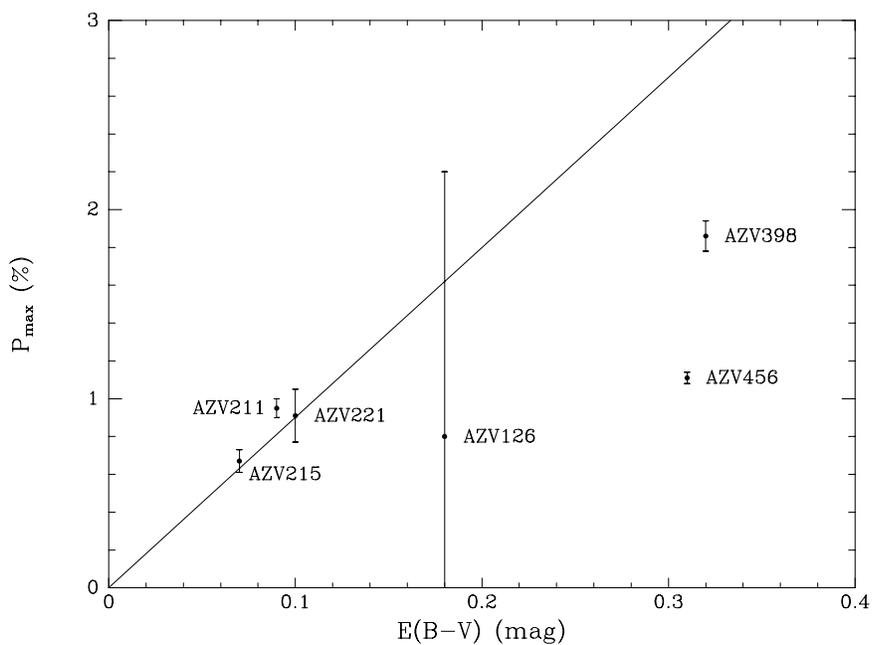,width=9.cm,angle=-90}
\caption{Polarization vs. $E(B-V)$. The continuous line represents
the Galactic upper limit. See text (sec. 3.2) for discussion.}
\label{pol_ebv}
\end{figure}

A plot of \pmax versus $E(B-V)$ for our sample is given in
Fig. \ref{pol_ebv}. We have used the 2-parameter fit
\pmax values from Table \ref{tab_serk}.
We have obtained the total color excesses
(Table \ref{tab_prog}, col. 7) using the spectral type-color relation by
Brunet (1975).
When both our estimates and other observed values
were available (AZV 398 and 456, Bouchet et al. 1985), we used the latter.
For AZV~215, we used the value of $+0.12^{mag}$ (Bouchet et al. 1985).
We have then corrected all color excesses by $0.05^{mag}$ to take into
account the Galactic foreground reddening (sec. \ref{for_pol}; Bessel 1991).
The above empirical relation between \pmax and color excess for the Galaxy is
the solid line plotted in Fig. \ref{pol_ebv}.
It is seen that the SMC stars also obey the Galactic relation between
\pmax and $E(B-V)$.

A related quantity of interest is the average ratio \ppmax/$E(B-V)$ for
the SMC objects. This is 7.2 \%$mag^{-1}$ (from Tables \ref{tab_prog}
and \ref{tab_serk}). This value is comparable to the corresponding
values for the Galaxy (5.0 \%$mag^{-1}$; Serkowski et al. 1975) and the
LMC (6.0 \%$mag^{-1}$; Clayton \& Martin 1985). Our average is biased
due to our selection of the more highly polarized stars for this
multiwavelength study. Data
from our survey in progress should be able in the near future to
improve the estimate of this ratio for the SMC.

     Serkowski et al. (1975) found a relationship between \lmax
and R (=A(V)/$E(B-V)$), R = 5.5 \llmax, for stars along several
lines of sight in the Galaxy. Whittet \& van Breda (1978), with
the aid of infrared photometry, confirmed that relation and
obtained R = (5.6$\pm$0.3)\llmax. This correlation was re-examined
and again confirmed by Clayton \& Mathis (1988) using sight
lines which included dense clouds as well as the more diffuse
ISM. Using the data for AZV~211, 221 and 398, we obtain
$\langle$\llmax $\rangle$=(0.40$\pm$0.02)\mic. Bouchet et al.
(1985) obtained from visual and near IR photometry for stars in
the SMC a value of R = 2.72$\pm$0.21, from which we obtain R/\lmax
= (6.8$\pm$0.6), still consistent with the Galactic relation. In
other words, the somewhat smaller value of R for the SMC does
translate into smaller values of \lmax. More SMC
data are clearly needed to examine this relation further.

For AZV~456, the \lmax
values from the 2- and 3-parameter fits (Table \ref{tab_serk}, col. 2) give,
using the above Galactic relation between R and \lmax,
3.30$\pm$.24 and 3.19$\pm$.20 for R. In other words, the star
with a 'Galactic' extinction curve does suggest a larger value
of R than that inferred from the typical SMC extinction.
We have however already pointed out in sec. \ref{serk_fit}
that AZV~456 does not seem to conform to the Galactic K vs. \lmax.
In addition, we shall see in sec. \ref{param} that neither does its extinction
curve conform in detail to that expected from a larger value
of R.

\section{Parametric study of UV extinction}
\subsection{Parametrization}
\label{param}

     In order to analyse objectively the extinction curves for
the SMC we have for the first time fit them by using the
parametrization of Fitzpatrick \& Massa (1986). However, we have
fit the parameters simultaneously, contrary to the approach of
Fitzpatrick \& Massa (1990). We have done this by minimizing the
chi-square. The parametrization of Fitzpatrick \& Massa (1990)
is expressed by

\begin{equation}
\label{eq_fm}
\frac{E(x-V)}{E(B-V)}=c_1+c_2 x + c_3 D(x;\gamma,x_o) + c_4F(x)\
,
\end{equation}

\noindent where
$$ x= \lambda^{-1}\ ,$$
$$ D(x;\gamma,x_o) = \frac{x^2}{(x^2-x^2_o)^2 + x^2\gamma^2} $$

\noindent and

\[ F(x) = \left\{ \begin{array}{lc}
0.5392 (x-5.9)^2 + 0.05644(x-5.9)^3, & if\ x \ge 5.9\micron^{-1};
\\
0.0\ , & if\ x < 5.9\micron^{-1}.
\end{array} \right. \]

     We have not considered the star AZV 20 because of its very
small signal-noise extinction curve (see Fig. \ref{fig_ext}a and discussion
in sec. \ref{ext_det}). We have performed the fits using the extinction
curves with bins of both 1 \AA~ (5 \AA~ in the case of AZV~126)
and of 80 \AA~\ to check the dependency of the parameters on the
bin size. The results are presented in Table \ref{tab_fm} and in
Fig. \ref{fig_fm}
together with the results from the fit of the Galactic curve
(Seaton 1979).

 We now discuss the fits for each star with reference to
Table \ref{tab_fm} and according to the three parts of equation \ref{eq_fm}:
the linear part, the Drude function fit to the bump $D(x;\gamma,x_o)$, and the
exponential fit to the increasing extinction into the FUV. Column 14
of Table \ref{tab_fm} gives the integral of the extinction curve over
the bump.

     AZV 211 and AZV 398 both show the typical SMC extinction
(Pr\'evot et al. 1984) with no bump and a rapid increase in the
UV and FUV extinction (Table \ref{tab_fm}, col. 8). Although these stars do
not present the bump, any oscillation in the extinction curve
might be interpreted artificially by the code as a small bump
(see Table \ref{tab_fm}, cols. 10-13 and Fig. \ref{fig_fm}). For that reason, we have
also performed the fits without the Drude function component.
There was however no significant difference in the resultant
parameters.

The extinction curve of AZV~456 is very
similar to that for the Galaxy (Fig. \ref{fig_fm}; Lequeux et al. 1984;
see entries for AZV~456 and the Galaxy in Table \ref{tab_fm}) and,
in fact, its extinction curve has been often referred to as a
'Galactic-type' curve. However, our fits show that its bump is
shifted to the blue ($x_o$ = [4.66$\pm$0.02]\clau, Table \ref{tab_fm}) with
respect to the Galactic average ($x_o$ = [4.596$\pm$0.019]\clau). For
comparison, the largest value of $x_o$ in the sample of Fitzpatrick
\& Massa (1986) is 4.63\clau. The width and intensity of the bump
for AZV 456 are within the range of those for the Galaxy. In
contrast, the three sight lines which show $x_o$ $>$ 4.65\clau, seen
through dense material, all appear to be associated with broad
bumps (Cardelli \& Savage 1988; Cardelli \& Clayton 1991; Mathis
1994).

\begin{figure}
\epsfig{file=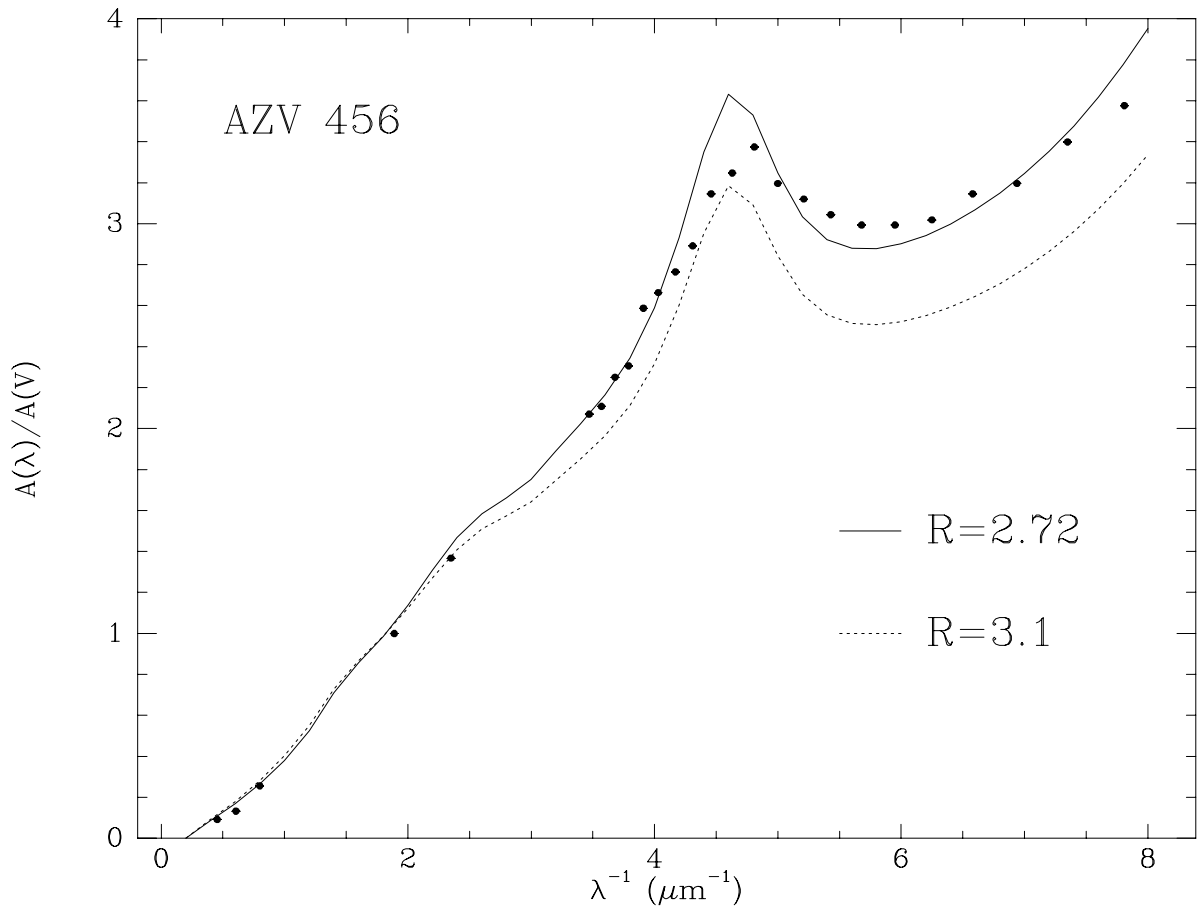,width=13cm,angle=-90}
\caption{The extinction curve of AZV~456 with the analytical
curves of Cardelli et al. (1989) for R=2.72 (solid line) and
R=3.1 (dashed line) superimposed.}
\label{fig_ccm}
\end{figure}

     From studies of several Galactic sight lines, which included
diffuse, dark cloud and star formation regions, Cardelli, Clayton
\& Mathis (1989) have shown that there is an average extinction
law over the wavelength range 3.5\mmic to 0.125\mmic which is
applicable to such environments. This mean extinction law,
A($\lambda$)/A(V), depends only on the parameter R. They have
noted, however, that a few sight lines, which included ones
with broad bumps and those towards the LMC, did not conform to
such a law. In Fig. \ref{fig_ccm} we plot the UV extinction curve of AZV~456
with the analytical law of Cardelli et al. (1989) for the values
of R = 2.72, 3.1. It can be seen that the observed and analytical
curves are discrepant, specially around the bump region, meaning
that the line of sight to AZV~456 does not conform to the single
parameter interpolation that performs well for the Galactic
environs.

     Among stars in our Galaxy which also present non-standard UV
bumps, HD 62542 has the most extreme bump, centered at
4.74\ccla (Cardelli \& Savage 1988). Interestingly enough, the
FUV extinction curve of HD 62542 and the SMC are actually very similar.
However, its bump is extremely shallow and its intensity is
quite different (lower) compared to the Galactic average and
AZV~456. In addition, our multicolor polarimetry of HD 62542
(reported by Clayton et al. 1992) shows that its \lmax value is
0.59\mmic, quite different from the low \lmax values for the SMC
(Table \ref{tab_serk}). This stresses the value of linear polarimetry in
providing additional, independent information about the grains.
In fact, in sec. \ref{models} it will become clear that detailed
simultaneous fitting of both extinction and polarization is more
restrictive on the grain model parameters.

     The parametrization of the extinction curve for AZV 126 with 1
\AA~ bins did not converge. This may have been due to the available
signal-to-noise or some spurious effect in the extinction curve that
the parametrization fit tried to include.  We have hence fit eq. \ref{eq_fm} to
an extinction curve binned to 5 \AA\ for AZV 126. This allowed a more
reasonable parametrization.  It nevertheless shows an abnormally small
slope for the linear part (Table \ref{tab_fm}, col. 4). The Drude
function fits (Table \ref{tab_fm}, cols.  10-13) indicate that it may
have the extinction bump but shifted to the blue (Table \ref{tab_fm},
col. 12) with regards to both the Galaxy and AZV~456. The bump is also
abnormally wide (Table \ref{tab_fm}, col. 10), although its intensity
is within the Galactic range. The rise into the UV and FUV (Table
\ref{tab_fm}, col. 8) is intermediate between that typical for the SMC
and that for the Galaxy. The extinction curve for AZV 126 has a large
intrinsic uncertainty and it must be viewed with caution.

\subsection{Correlations between extinction components}
\label{corr_ext}

\begin{figure}
\epsfig{file=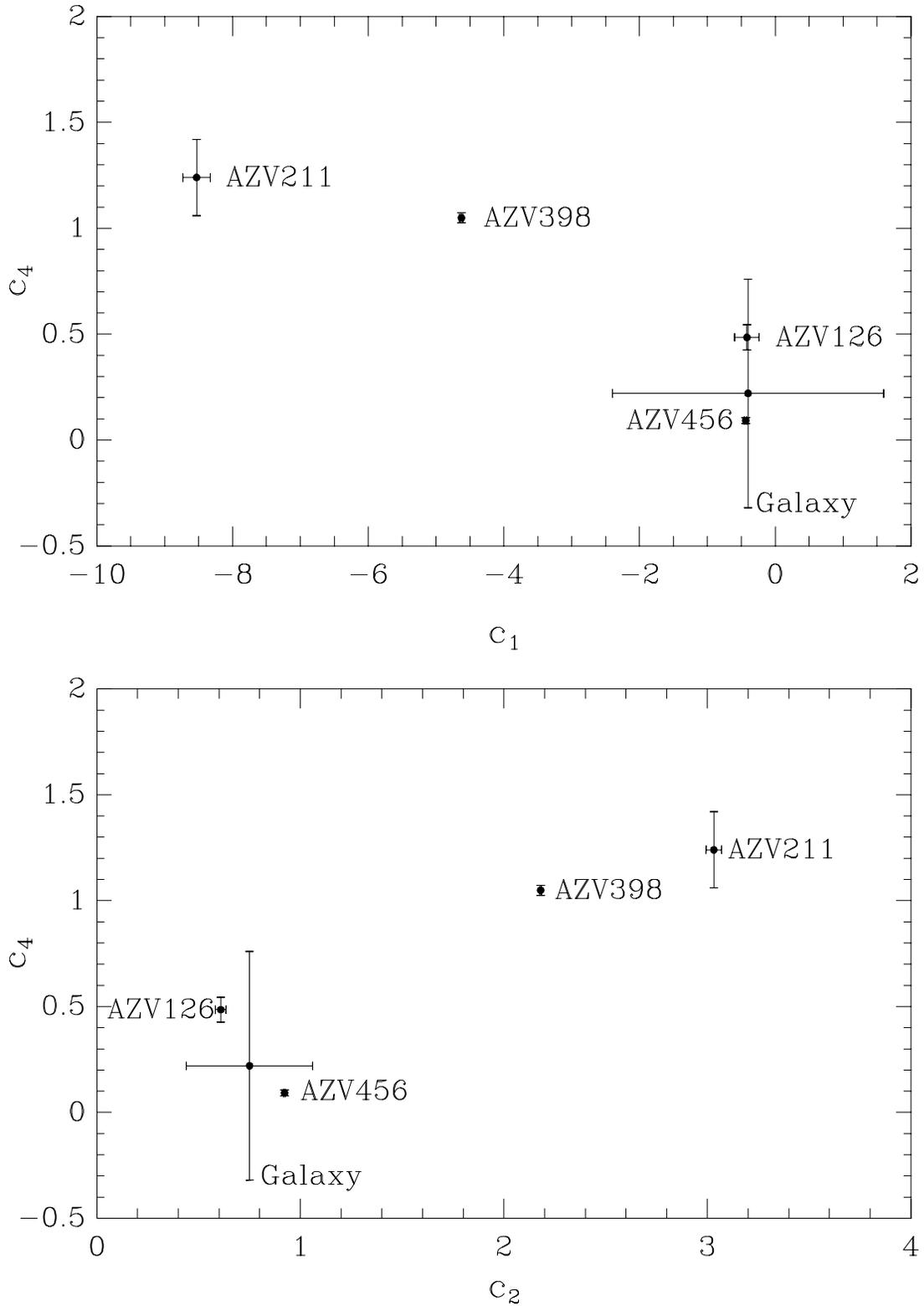,width=16cm}
\label{fig_cof}
\caption{The coefficient, $c_4$, for the exponential UV and FUV
parts of the extinction curve plotted against the coefficients,
$c_1$ and $c_2$, respectively, for the linear part of the extinction
curve. See eq. 5.}
\end{figure}

     Although the small number of extinction curves available for
the SMC makes any search for correlations between the parameters
difficult, there are tentative indications of correlations 
between the linear coefficients ($c_1$, $c_2$) and the coefficient
($c_4$) of the linear and FUV portions, respectively, of the
extinction curves (eq. \ref{eq_fm}). These are shown in Fig.
\ref{fig_cof}. The point
with large error bars in that figure corresponds to parameters
for the Galactic curve given in Table \ref{tab_fm}. The large errors are due
to the fact that relatively few data points were conveniently tabulated for
the fit to the Galactic curve. Those error bars are hence very conservative.

Figure \ref{fig_cof} 
shows that the FUV curvature for the SMC, $c_4$, is anti-correlated
with the constant component, $c_1$, and positively correlated with the
slope of the linear component, $c_2$. In fact, the parameters $c_1$ and $c_2$
are perfectly
correlated (Fitzpatrick \& Massa 1988; Jenniskens \& Greenberg 1993) and
$c_1$ is not an independent parameter.
If the FUV curvature does increase
at the same time as the contribution of the linear component to the extinction,
this can be interpreted as the result of a simultaneous decrease in the 
average size of the grains responsible for these various parts of
the extinction curve. For the Galaxy no correlation is found
between the linear rise and the extinction in the FUV (Jenniskens \&
Greenberg 1993). This may
be a further indication of the differences between the ISM in the 
SMC and the Galaxy, where the correlation found may simply signify
the increasingly larger number of smaller particles as we go from the
Galaxy to the SMC.

The fact that AZV~211 and AZV~398 show a sizeable contribution to
the FUV term in their extinction curves, quite larger than that of
AZV~456, is consistent with what is found for the Galaxy, for which no
positive correlation exists
between the bump and the FUV curvature (Greenberg \& Chlewicki
1983; Jenniskens \& Greenberg 1993). In the Galaxy, the linear rise part
($c_1$, $c_2$) of the extinction curve
is systematically less in dense media (Cardelli et al. 1989;
Jenniskens \& Greenberg 1993). The behavior of the extinction curve for AZV~456
seems to follow the same trend. In contrast, the Galactic object
HD204827 shows a steep FUV rise (Fitzpatrick \& Massa 1990) while at the
same time showing small \lmax (Whittet et al. 1992; Clayton et al. 1995),
in line with the typical SMC sight lines. In section \ref{discussion},
we will return to this discussion.

\section{The SMC Data and Dust Models}
\label{models}

     We have tried to fit the SMC data using the MRN grain model
(Mathis, Rumpl \& Nordsieck 1977).  In this model the grains are
homogeneous spherical particles of silicate and of graphite with
a -3.5 power law size distribution. The sizes range from
0.02\mmic to 0.25\mmic for silicate grains and from 0.005\mmic to
0.25\mmic for graphite grains. There are two approaches to the
polarization within the context of the MRN model. Mathis (1979,
1986) introduced an additional population of elongated silicate
grains (cylinders). A single size distribution describes both the
spherical and the elongated silicates. From a minimum grain
radius, $a^{sil}_-$, until an intermediate radius, $a^{sil}_p$, the
grains were assumed to be spherical (or not aligned), and from
then up to a maximum radius, $a^{sil}_+$, they were taken as
aligned cylinders. This case will be called the M79 model. 
Mathis (1986, hereafter M86) proposed a modification to the above
scenario. The polarizing material would consist of silicate
particles containing inclusions of ferromagnetic material in
order to make the alignment more efficient (Jones \& Spitzer
1967). The number of these inclusions increases with grain size
in such wise that larger grains have a greater probability of
being aligned. Although the size distribution is the same as in
M79 model, in this case there is a probability function which
must be multiplied by the size distribution of cylinders in order
to give the number of aligned cylindrical grains contributing to
the polarization. To calculate the extinction in the M86 model,
we have assumed that all cylindrical grains are aligned.

     We introduce two innovations to the above models. First, we
attempt to fit the observations by using the polarization and the
extinction together. The implications on the extinction curve
of including silicate cylinders has not been studied in detail by
Mathis (1979, 1986). Kim \& Martin (1994, 1995) have derived the
dust particle mass distribution from the polarization curve for
the Galaxy, although, as noted by them, extinction and polarization
curves still need to be consistently and simultaneously interpreted.
Thus, to the best of our knowledge such combined
fits have not been attempted previously. To carry out these fits, 
it is necessary for us to adopt
a model for the degree of alignment as a function of grain size, and an
axial ratio for the aligned grains. At the end of the fitting process we
look at the predicted ratio P(V)/A(V) to check the validity of these
assumptions. It is important to realize (sec. \ref{galaxy}) that
the combination of shapes may affect
the extinction in addition to the polarization. Secondly, we
introduce {\it volume} continuity as an alternative to {\it size} continuity. 
The size distribution, n(a), can be algebraically represented as

n(a) = N f(a),

\noindent where a is the particle radius and f(a) describes the shape of
the size distribution. N, called hereafter number constant, is
related to the absolute number of grains. This constant is
dependent on the elemental abundance (gas phase plus dust) and on
the depletion of the main grain constituents. We use the word
depletion to mean the fraction of a given chemical element locked
up in the grains. The abundances of interest are those of carbon
(C) and silicon (Si).

     In the single size distribution of Mathis (1979), when
silicate cylinders and spheres have the same number constant, the
Si fraction in each population is automatically fixed by one
boundary condition: the number of cylinders and spheres at the
radius $a^{sil}_p$ must be the same. Hereafter, we call this case
{\it size continuity}. We have in addition studied the influence of a
different boundary condition on the distribution of spheres and
cylinders. We employ the same shape, {\it f(a)}, for the size
distributions of both populations, but their number constants are
different. Specifically, we have calculated them in such a way
that the volume distribution is continuous, i.e., the boundary
condition is such that the total volume occupied by the spherical
and by the cylindrical grains of size $a^{sil}_p$ must be equal.
This will be called {\it volume continuity}. 
For example, for the elongation of two adopted here
for the cylindrical particles, the number of spheres is larger by a
factor three. This results in a larger
relative contribution by spherical particles in the case of the
volume continuity as compared to size continuity.

     The optical properties for the so-called astronomical
silicate have been taken from Draine \& Lee (1984), who made a
synthesis of laboratory and astronomical data. Its properties are
quite similar to olivine ($[Mg,Fe]_2SiO_4$) and it shows the
characteristic change in the UV slope of the extinction curve
around 6.5\clau. For enstatite ($[Mg,Fe]SiO_3$), we have used the
optical constants obtained by Huffman \& Stapp (1971).
These constants are given only for the optical and UV. This
material was employed by Bromage \& Nandy (1983) in their fit to
the typical extinction curve for the SMC. The index of refraction
for graphite has been taken from Draine \& Lee (1984). We use the
"1/3-2/3" approximation to calculate the extinction coefficients.
Draine \& Malhotra (1993) have recently shown that this procedure
is sufficiently accurate. For the amorphous carbon we have
adopted the constants of Duley (1984).
     	
\subsection{Fit procedure}
	
     Our approach was to first fit the polarization,
the wavelength dependence of which determines the size
distribution parameters of the cylindrical silicate population.
Then from the maximum polarization of the fit we determine the Si
abundance in the form of cylindrical grains. Next we fit the size
distribution parameters of graphite and spherical silicate grains
in order to reproduce the extinction curve, assuming the
cylindrical population fixed.

     The minimum and maximum sizes of the cylinder silicate
distribution were obtained from a chi-square (\qqui) minimization
procedure applied to the polarization curve normalized to the
maximum polarization. We assumed perfect spinning alignment (Chlewicki \&
Greenberg 1990 and references therein) and three values of the
angle between the direction of the magnetic field and the plane
of sky, $\gamma$: 10, 30 and 60 degrees. Next we calculated the
necessary Si abundance (i.e., Si/H) to reproduce the polarization in the V
filter. For that we must know the color excess, $E(B-V)$, and the
gas to dust ratio, N(H)/$E(B-V)$ (e.g., Casey 1991), since we are
in effect fitting P(V)/N(H). Assuming a given continuity (size
or volume), we can calculate the number constant of the spherical
silicate distribution, whose maximum size is fixed by the
cylinder sizes.  Therefore, for a given C depletion, we have only
to fit the minimum size of spherical silicates and the size
parameters of the C grain population (again the slope of
the size distribution was held fixed, for simplicity).
We also apply a chi-square
minimization procedure to the A(\lamb)/N(H) data.  All the extinction
curves were fit using both size and volume distribution
continuities and various carbon depletions.

\subsection{Fits to Galactic Data}
\label{galaxy}

\begin{figure}
\epsfig{file=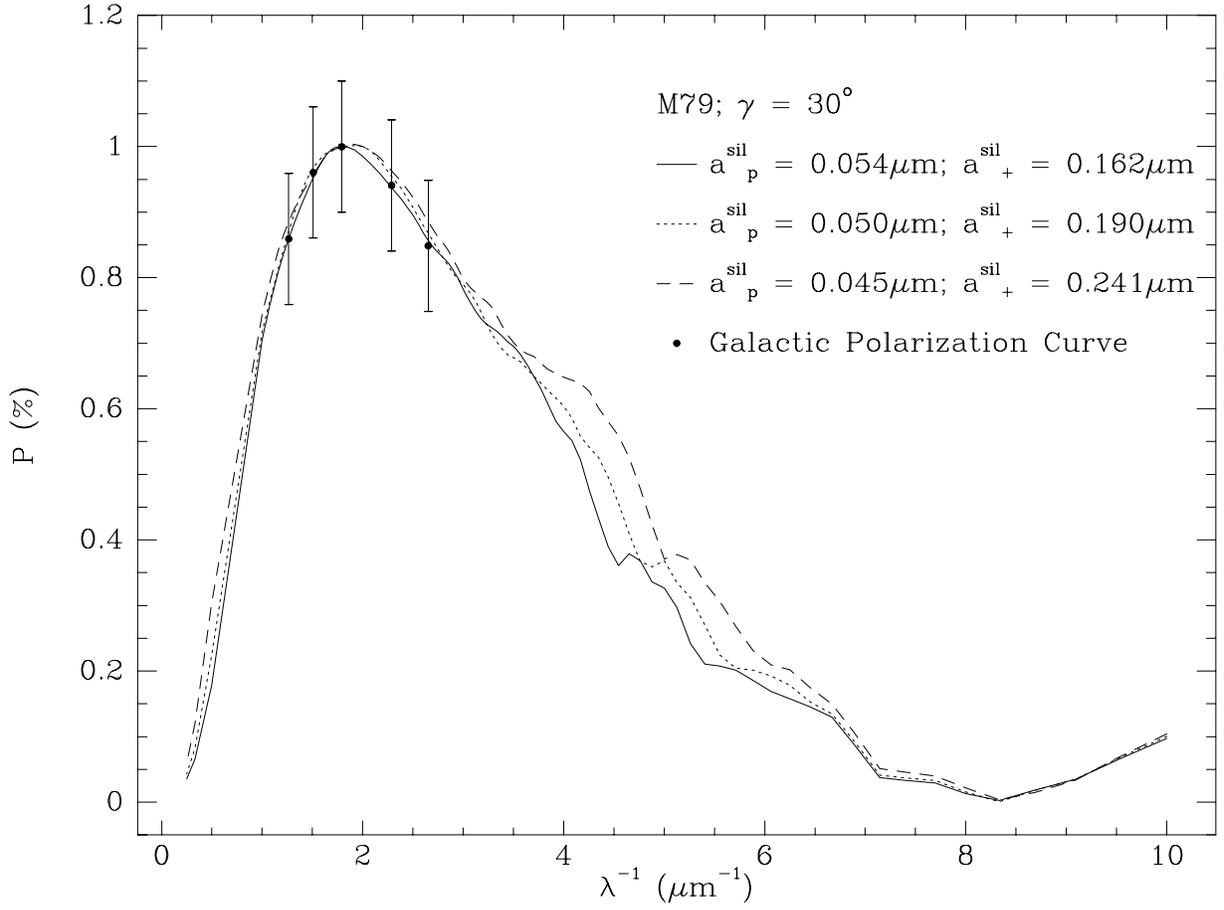,width=13cm,angle=-90}
\caption{Fits to the Galactic polarization curve for $\gamma$ =
30$^o$. The different curves are for size parameters representing
local \qui minima (Table 6).}
\label{gal_pol}
\end{figure}

     In order to compare the SMC environment with that of the
Galaxy we first discuss fits to the polarization and extinction
in the latter. In Fig. \ref{gal_pol} and Table \ref{aj_pol}
we present fits to the polarization curve which
was considered as a Serkowski law with \lmax = 0.55\mmic from
optical studies. Presently, the interstellar polarization curve
is known throughout a much wider wavelength range, which could
provide stronger constraints on the size distribution (e.g., Kim
\& Martin 1994; Anderson et al. 1996). However we decided to
consider only the optical range, which is the one we have available
for the SMC (sec. \ref{polarization_data}).
To obtain the value of P(V)/N(H) to fit, we adopted \ppmax/$E(B-V)$ = 9 (the
maximum observed, and P(V) $\approx$ \pmax), 
N(H)/$E(B-V)$ = $5.848\ 10^{21}$ cm$^{-2}$ (Bohlin, Savage \& Drake 1978), and
R = 3.1. The fractional error in P(\lamb) was taken to be 10\%.  In
Figure 9 the scale of P is in fact arbitrary, corresponding to
N(H) = $6.433\ 10^{20}$ cm$^{-2}$, or $E(B-V)$ = 0.11.
We have assumed the following abundances:
Si/H = $3.55\ 10^{-5}$ (7.55 dex) and C/H = $3.63\ 10^{-4}$ (8.56 dex)
(Anders \& Grevesse 1989). In
Fig. \ref{gal_pol} we show fits for $\gamma$= $30^o$, although the wavelength
dependence of the Galactic polarization curve is well fit for all
$\gamma$ values used. We found local chi-square minima for the various
size distributions listed in Fig. \ref{gal_pol}. Below we will study the
behavior of the extinction curve for these different cylinder
size distributions. The different size parameters obtained for
a given $\gamma$ differ in range, but have practically the same
$\langle a\rangle$ (Table \ref{aj_pol}). For $\gamma$ = $60^o$ the 
required Si abundance is
greater than that available in the Galactic ISM. In this case we have assumed a
\ppmax/$E(B-V)$ of 2.0 and 5.0 (Table \ref{aj_ext}, col. 16). This
lowers the amount of Si needed to fit the polarization.

\begin{figure}
\epsfig{file=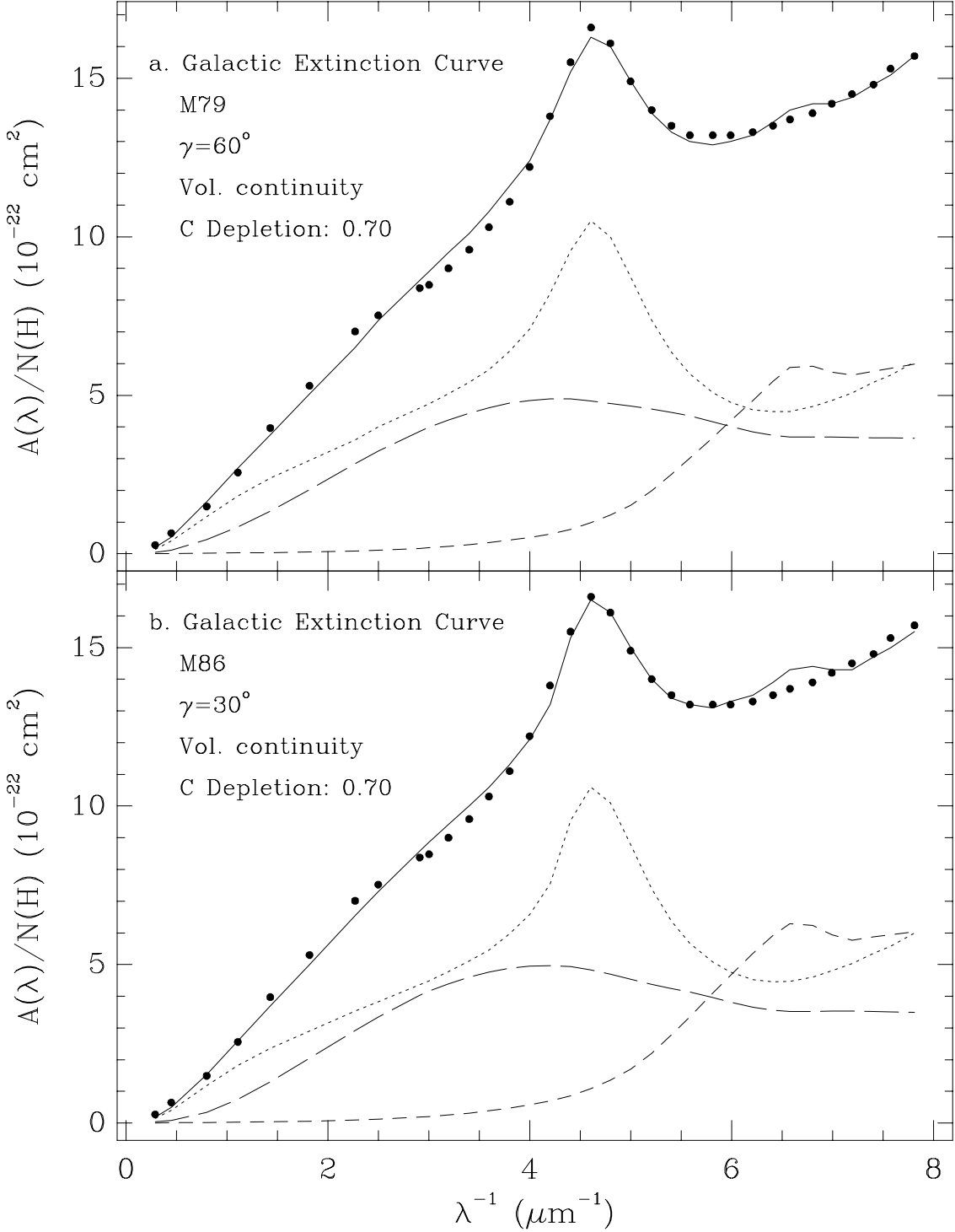,width=16cm}
\caption{Best fits to the Galactic extinction curve (solid dots)
using cylinder sizes derived from the M79 (a) or M86 (b) model
fits. Additional fit parameters are given in Table 7.
The dotted line represents the contribution of graphite;
the long dashed line, the contribution of silicate cylinders; the
short dashed line, the contribution of silicate spheres and the
solid line the total extinction.}
\label{gal_ext}
\end{figure}

We now proceed with fits to the Galactic extinction curve.
These are shown in Fig. \ref{gal_ext} and Table \ref{aj_ext}.
We see from Table \ref{aj_ext} that
the best fit to the Galactic extinction curve for each cylinder
parameter set is systematically found for a Carbon depletion of about
0.70 and volume continuity (Fig. \ref{gal_ext}). The curves depart
from the observed Galactic curve in the optical range and this is
due to the adopted size distribution, as demonstrated by Kim,
Martin \& Hendry (1994). The better agreement of the extinction
curve using volume continuity would seem to favor a situation
in which the smaller grains are more numerous than predicted
by the power law of index -3.5. This may be accomplished by
a steeper power law or a discontinuous size distribution
from spheres to cylinders, as we assumed.

In any case, an important point is that, if we start with the size
distribution of spherical silicates that, together with graphite,
fit the extinction curve, a simple replacement of the larger silicate
particles by cylinders that fit P(\lamb) will {\it not} fit the
extinction any longer.

A glance through the fits for the Galaxy in Table \ref{aj_ext} reveals
that there is a tendency of both the M79 and M86 models with perfect
spinning alignment to provide relatively high P(V)/A(V) values. This
would indicate that perfect spinning alignment is not a necessity (see also
Kim \& Martin 1995).

\subsection{SMC fits}

We have performed polarization and extinction fits for two
stars in the SMC:  AZV~398 and AZV~456, since they have
relatively high values of polarization and better determined
extinction curves. Furthermore, they may represent lines of sight
with different grain characteristics. Although their UV
extinction curves are different, their IR extinction
seems to be very similar according to Bouchet et al. (1985). We
have hence used the value of R (= 2.72) from the latter reference. The
abundances of carbon (C/H = 6.455\ 10$^{-5}$, or 7.80 dex) and silicon
(Si/H = 7.59\ 10$^{-6}$, or 6.88 dex)
have been taken from Dufton, 
Fitzsimmons \& Howarth (1990), which provides the abundances of C, Si 
and other elements in the atmosphere of a main sequence B-type star 
in the SMC. We assume that these represent the present ISM abundance 
in that galaxy (Pagel 1993). For our subsequent analysis, we
assumed the following N(HI)/A(V) values: for AZV~398, 4.1\ 10$^{22}$ cm$^{-2}$;
for AZV~456, 6.9\ 10$^{21}$ cm$^{-2}$ (Bouchet et al. 1985).

\subsubsection{AZV~398}
\label{av398}

\begin{figure}
\epsfig{file=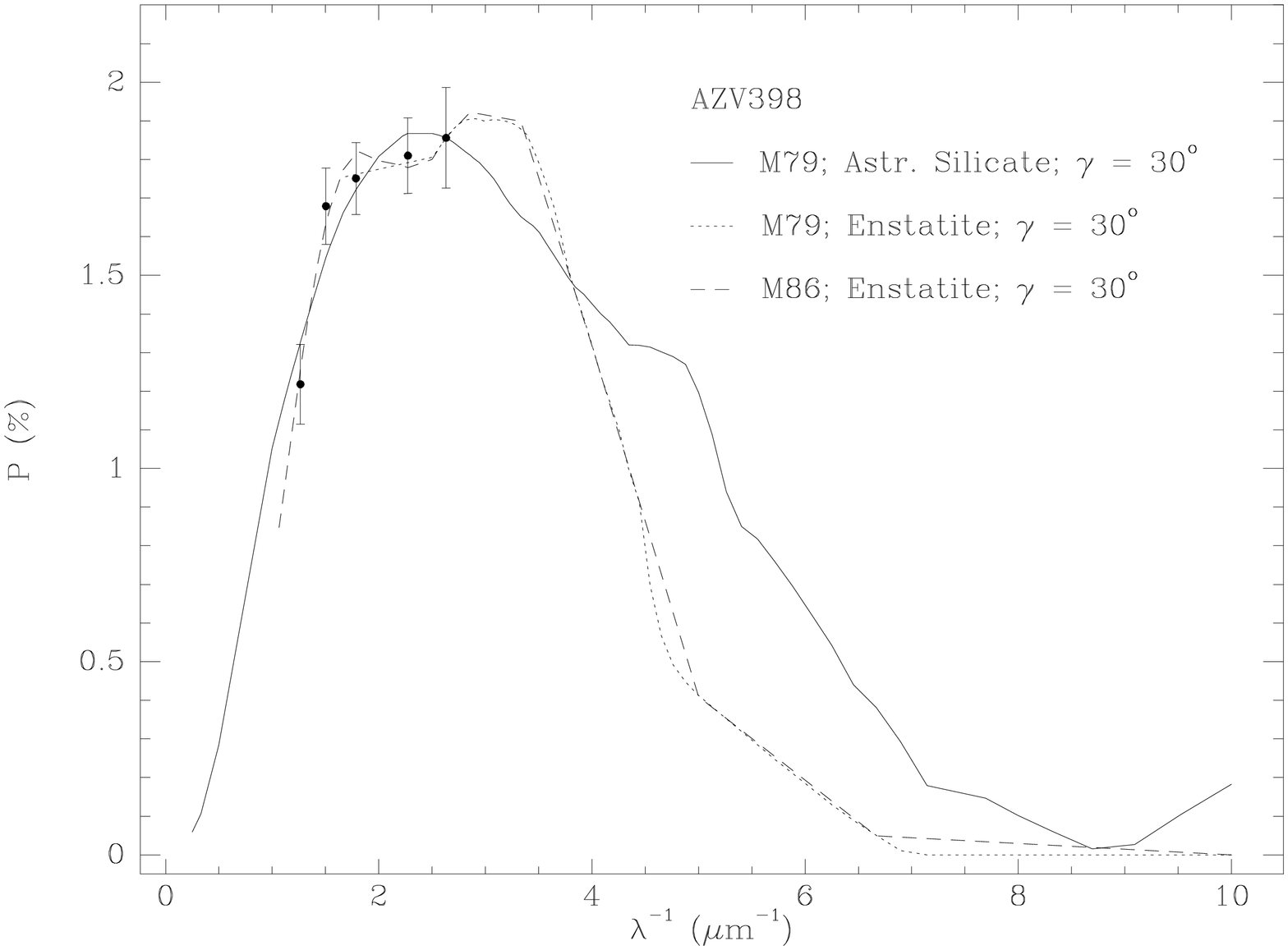,width=13cm,angle=-90}
\caption{Fits to the AZV~398 polarization curve. Fit parameters are given in
Table 6.}
\label{398_pol}
\end{figure}

    In Fig. \ref{398_pol} and Table \ref{aj_pol} we present 
the model fits to the
polarization of AZV 398 which shows the highest polarization in
our SMC sample (sec. \ref{polarization_data}). The polarization decreases
rapidly from the R to the I
filter and not all of the models fit those points equally well.
We also performed the fits by replacing
astronomical silicate with enstatite. The value of $\langle a\rangle$ obtained
with enstatite is slightly larger than that obtained using
astronomical silicate as a consequence of the differences in
optical constants of those materials.

\begin{figure}
\epsfig{file=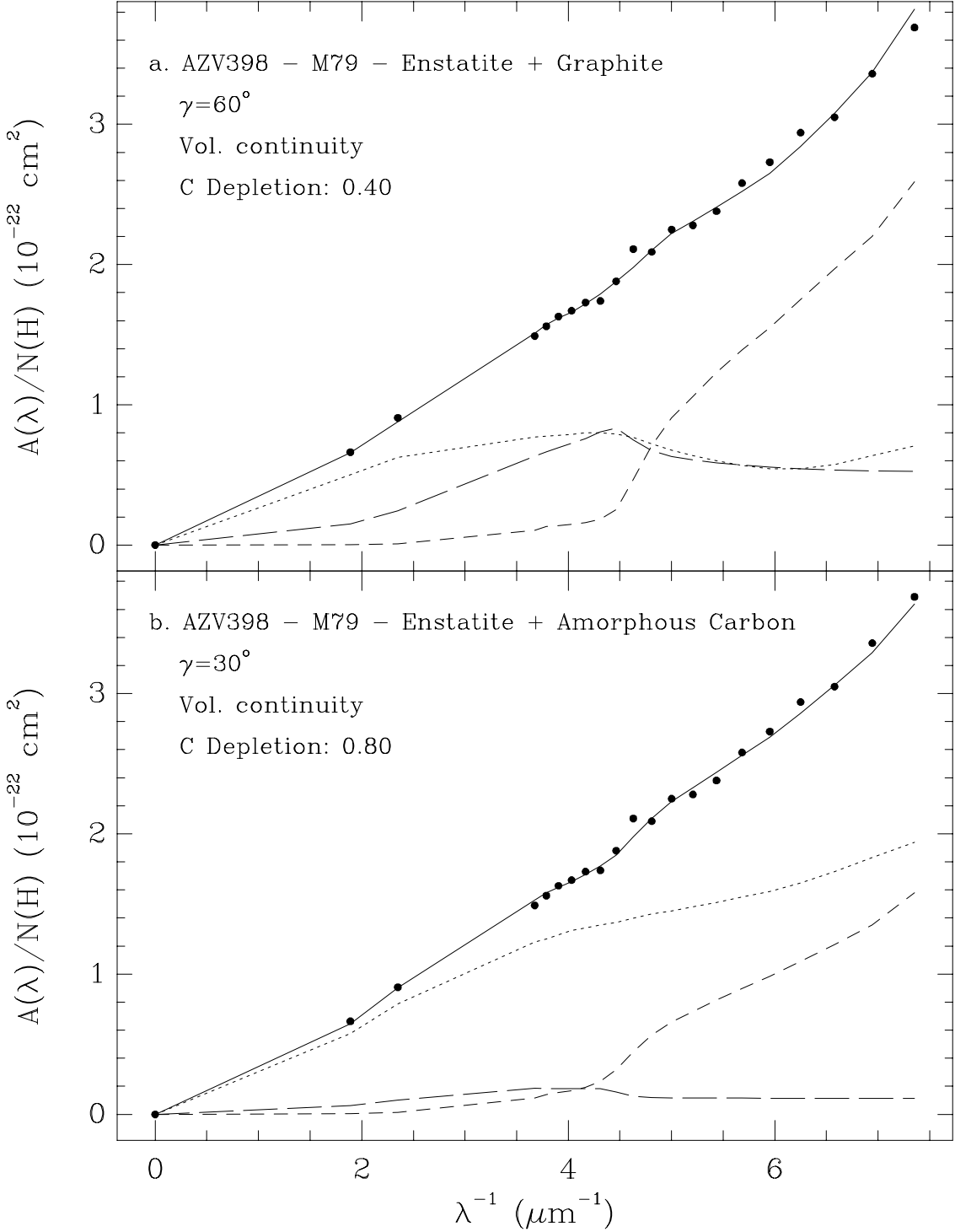,width=16cm}
\caption{Best fits to the AZV~398 extinction curve using graphite (a) or
amorphous carbon (b). Fit parameters are given in
Table 7. Line styles are as in Fig. 9.}
\label{398_ext}
\end{figure}

     In Fig. \ref{398_ext} and Table \ref{aj_ext} we present
the model fits to the extinction of
AZV 398. We obtain a good fit by using enstatite plus graphite
and volume continuity (Table \ref{aj_ext} and Fig. 
\ref{398_ext}a). In this case,
the extinction curves of each population can be combined to give
a practically linear curve. 40\% of the available C in the SMC is used.
This fit, however, requires
approximately 30\% more Si than the available in the SMC ISM (Table 
\ref{aj_ext}, col. 14). Also of note is the relatively high value
of the lower limit of the graphite size distribution (0.0372 \mic, Table
\ref{aj_ext}) of this fit, which prevents the extinction curve to show
an UV bump.

     We have examined the consequences of replacing the
population of graphite grains by one of amorphous carbon
particles.  As seen in Fig. \ref{398_ext}b we can reproduce the AZV~398
extinction curve in shape and level. The amorphous carbon
extinction fits specially well the FUV and because of that the fits
with large C depletion tend to have smaller \qqui. Also, we have obtained
a larger number of good fits with amorphous carbon
compared with fits using graphite. The required
amount of Si and C in grains is about 70\% and 80\%, respectively,
of that available in the SMC (Table \ref{aj_ext}).

The required mass in solid carbon obtained from the AZV~398 fits
(in boldface in Table \ref{aj_ext}; Fig. \ref{398_ext}) varies from 10\%
(with graphite) to 20\% (with amorphous carbon) of the corresponding value
derived from the fits to the Galaxy. They reflect the product
(C-abundance)*(required depletion) from the Galaxy to the SMC. As
the required depletions are similar from one galaxy to another, that value
follows basically the C-abundance ratio between the two galaxies.
The same occurs
with the total grain mass (carbonaceous + silicate grains). However, in the SMC
the fraction of mass in cylindrical grains relative to the
total silicate grain mass (38\% with graphite and 22\% with amorphous carbon)
is significantly smaller compared to the corresponding value in the Galaxy
(about 70\%). Also, the upper limit of the carbon particle size distribution
(Table \ref{aj_ext}, col. 10) is significantly smaller than that in the
Galaxy for most of the
(and certainly for the best) SMC fits. The amorphous carbon size distribution
is also narrower than that for the Galaxy.
In other words, the dust particles are on average smaller in the SMC as
compared to the Galaxy. At the same time, note the absence of the very small
graphite particles that produce an extinction bump.

Another important corollary of these AZV398 fits is that the typical SMC
extinction curve cannot be adequately fitted by using silicates alone,
as previously advocated (Bromage \& Nandy 1983; Pei 1992), unless
the Si abundance in the SMC ISM is significantly revised upwards, an
unlikely event. Therefore,
we feel that use of the dust properties of models that employ silicates only
should be exercised with caution.

\subsubsection{AZV456}
\label{av456}

\begin{figure}
\epsfig{file=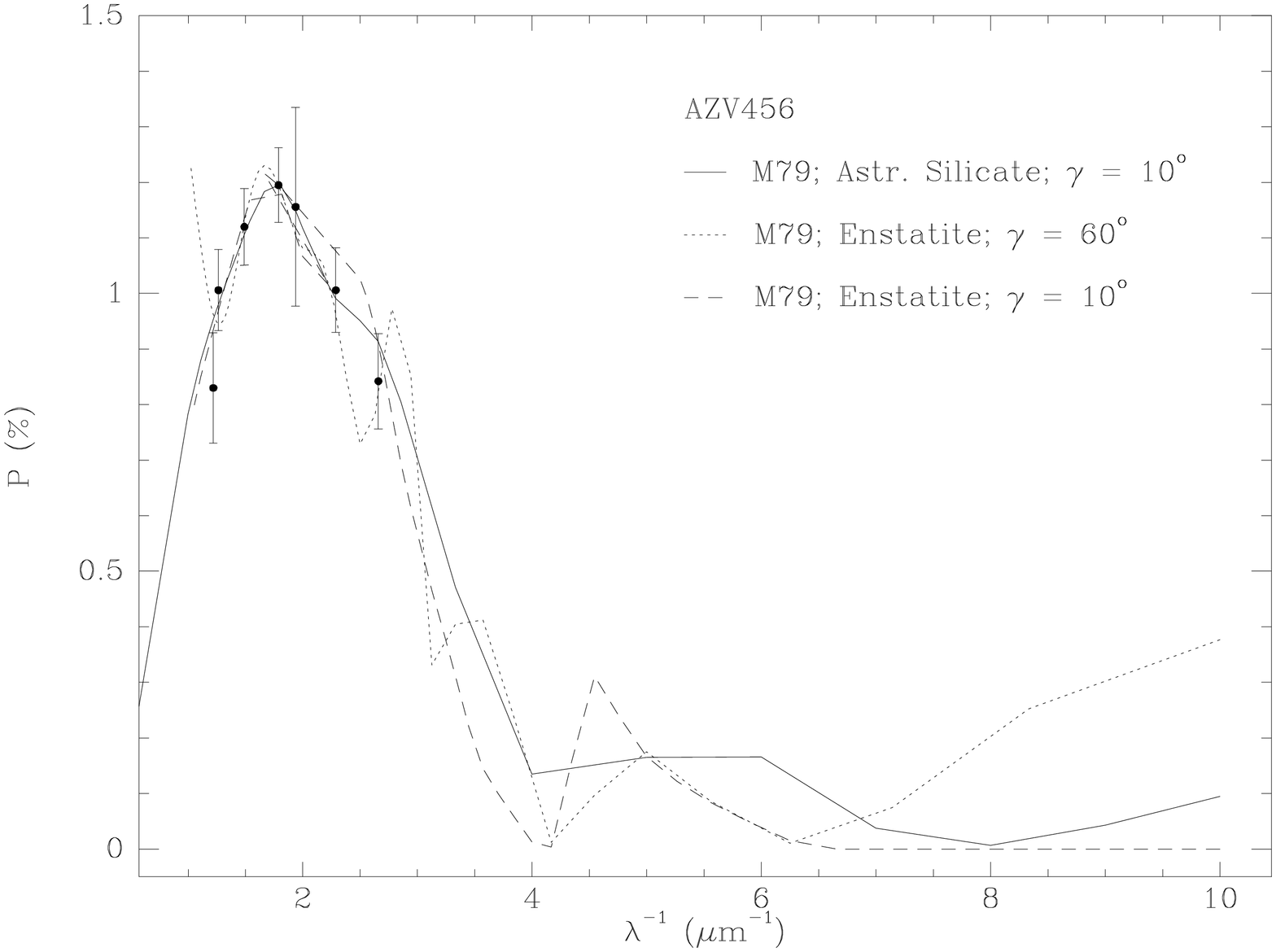,width=9cm,angle=-90}
\caption{Fits to the AZV~456 polarization curve. Fit parameters are given in
Table 6.}
\label{456_pol}
\end{figure}

In Fig. \ref{456_pol} and in Table \ref{aj_pol} we present
some of the best fits
obtained for AZV456 which has an unusually narrow polarization
curve. These occur for either small values of $\gamma$ or for
relatively narrow size distributions. The M79 fit with $\gamma$=$10^o$
and astronomical silicates (solid line) presents a significant improvement
relative to the
$\gamma$=$60^o$ fit with enstatite (dotted line). At the same time, the low $\gamma$
fits require less Si and provide less extinction (see below).
Values of $\gamma$ smaller than $10^o$
were also tried but did not improve the fits. Some curves from
models with narrow size distributions present an oscillating
spectral dependence, especially for larger
values of $\gamma$. The polarization curve of a single grain is
characterized by ripples which are usually averaged out by a size
distribution, except when the distribution is very narrow.

These fits are characterized by average sizes similar to 
those for the Galaxy, confirming the usual relationship between
\lmax and the average sizes. Wilking et al. (1980) suggested that
an increase in the real part of the index of refraction can
produce progressively narrower polarization curves. We have made
some fits using hypothetical compounds, with a
wavelength-independent index of refraction, in order to gain 
some insight into this. We have varied the real part of the index
of refraction from 1.7 to 2.0 and the imaginary part from 0.01 to
0.05 but such changes in the index of refraction do not seem to
be large enough to improve the results. Spheroids provide
narrower polarization curves than do infinite cylinders (Wolff et
al. 1994; Kim \& Martin 1995). Whether these particle types can
improve the fits remains to be verified.

Fitting the extinction of AZV~456 presents an interesting challenge,
given a Galactic gas-to-dust ratio and the low SMC metallicity.
We first try, as usual, to fit AZV~456
with silicate cylinders and spheres and with carbon spheres but,
because of the observed UV bump, we must now add graphite
spheres. The cylinders already use a sizeable fraction of the available Si
(Table \ref{aj_pol}, col. 6).
Providing the observed extinction levels by adding spherical particles
will require the addition of 
Si several times that available in the SMC. Graphite particles
reproduce the bump and provide some optical extinction, with a relatively
small UV contribution. It is
clear that there will not be enough material available in the SMC ISM
capable of providing the needed extinction towards AZV~456.

\begin{figure}
\epsfig{file=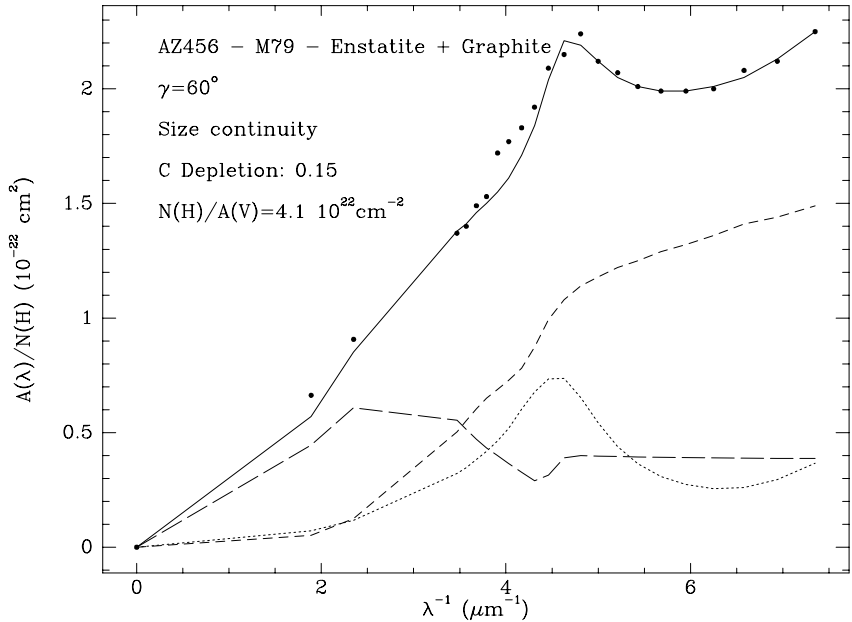,width=9cm,angle=-90}
\caption{The AZV~456 extinction curve, assuming that the gas-to-dust ratio
equals the N(HI)/A(V) ratio typical
of the SMC, that of AZV398. Dust parameters for the calculated solid line
are given in
Table 7. Line styles are: dotted = graphite; long-dashed = silicate cylinders
and short-dashed = silicate spheres.}
\label{456_ext_3}
\end{figure}

A possible solution out of this conundrum is to resort to the possibility
that AZV~456 may actually have a larger gas-to-dust ratio, closer to
the more typical SMC values. Lequeux (1994) has argued
for possible H$_{2}$ towards the AZV~456 line-of-sight
and we noted in the Introduction that there is an IRAS extended source
towards this same direction. Let us assume that the gas-to-dust ratio
towards AZV~456 equals the
N(HI)/A(V) ratio shown by AZV~398 (i.e., a factor of about 6 higher
than the observed N(HI)/A(V) ratio for AZV~456). For the cylindrical
grains, we take one of the grain populations that fits the polarization (Fig.
\ref{456_pol}), with a correspondingly lower Si/H abundance.
The resultant extinction fit is shown in Fig. \ref{456_ext_3}.
Note that the 'bluer' bump of AZV~456 is
now evident in Fig. \ref{456_ext_3}.
The cylinders use a relatively large amount (80\%) of the available
Si and provide a sizeable contribution to the extinction.
The total Si used is about 50\% more than the amount available. Our purpose
here is not to opt for a particular fit, given the
uncertain gas-to-dust ratio towards AZV~456, but only to show that it
may be possible to fit the extinction towards this star if we
assume a higher gas-to-dust ratio in its line-of-sight.

As a matter of fact, given that amorphous carbon (and silicates)
explains so well
the typical SMC extinction (sec. \ref{av398}), it is quite reasonable
to expect this component to be present towards AZV~456 as well. The
inclusion of amorphous carbon in a model will lower the amount of Si
needed, as this unpolarizing, carbon component will provide some of the needed
extinction as well as lower the P(V)/A(V) ratio. Additionally, the best
fits to the polarization of AZV~456
actually do use a small amount of Si (Table \ref{aj_pol}). 
The inclusion of amorphous carbon implies, of course, a larger number
of free parameters. We have however
used the same size distribution for amorphous carbon obtained
from the AZV~398 extinction fits and verified that the overall extinction
towards AZV~456 may indeed be obtained.

\section{Discussion}
\label{discussion}

\subsection{The SMC Interstellar Polarization and Extinction Data -
What do they mean?}

     We now wish to discuss the data presented on the sample of
stars in the SMC for which we have available both extinction
curves and/or wavelength dependent polarization data.

     Our extinction data suggest a correlation between the FUV
curvature and the linear portion of the extinction curve (sec.
4.2). The stars for which the FUV contribution is more important,
AZV 211 and AZV 398, do not show the bump, so the FUV curvature
cannot be attributed to the same grains producing the bump.
Further, these two stars show \lmax smaller than the Galactic
average. Other SMC stars also present small \lmax values (Table
\ref{tab_serk}, col. 2). Of the two stars in the SMC which have larger \lmax
values, similar to those for the Galaxy, the one with the best
signal-to-noise, AZV 456, has a well determined extinction curve
(Pr\'evot et al. 1984; sec. 4.1). This star does show the bump
and its \lmax value indicates that it is also characterized
by grains larger than those found for the other sight lines in 
the SMC. It is also of interest to note that from observations in our Galaxy
the bump is usually not polarized (Clayton et
al. 1992; Anderson et al. 1996), so that the grains producing the
bump are either spherical or not aligned.

It is tempting to conclude that the bump in the SMC would be
present only along lines of sight characterized by large \llmax.
This conclusion might be tenuous, since stars with both
polarization and extinction curves reliably determined are very
few in the SMC and there is only one known line of sight which
shows the bump. However, it is rather remarkable that AZV~456 has
an extinction curve similar to that for the Galaxy
(Pr\'evot et al. 1984; sec. 4.1) and a \lmax value similar to
the Galactic average (sec. 3.1). All of the remaining stars in
the sample of about 20 stars of Bouchet et al. (1985) and
Fitzpatrick (1985b) show a gas-to-dust ratio (N(HI)/A(V)) roughly 10 times the
Galactic average. The 3 stars in the sample of Pr\'evot et al.
(1984), which have this high gas-to-dust ratio, do not show the
bump. In addition, in our sample of stars those which do not show
the bump show small \lmax values. Extending the arguments of
Fitzpatrick (1989) to include polarization data, we may say that,
if the high gas-to-dust ratios are produced by conditions which
also give rise to the typical SMC extinction law and small \lmax
values, then our polarization measurements, as well as the
extinction law of Pr\'evot et al. (1984), may indeed represent the
typical extinction and polarization properties of the dust in the
SMC.

     The \lmax results from our polarization data are important in
order to improve the constraints on dust models for the SMC and
as a test for inferences based solely on extinction data. Bouchet
et al. (1985) point out that the value of R they have determined
for the SMC, slightly smaller than the Galactic value, might
suggest that the graphite grains would play a smaller role in the
visible/IR. According to them, the size distribution of silicate
grains, in the context of the MRN model would then have to be
shifted towards larger grains. Our fits to the observations show
that the opposite is actually true, \lmax is in general
smaller than in the Galaxy. While
it might be expected that \lmax would depend on how the alignment
mechanism works as a function of size, studies of the correlation
between extinction and polarization (Serkowski et al. 1975;
Whittet \& van Breda 1978; Clayton \& Mathis 1988; sec. 3.2)
for the Galaxy for several types of sight lines suggest that the
populations producing extinction and polarization would seem to be affected
similarly by the environment and that \lmax variations originate
mainly from differences in the size distributions rather than
variations in the alignment of particles (Clayton \& Mathis
1988).

     The \lmax results can be interpreted in terms of the dust grain
sizes. Values of \lmax close to 0.55\mmic indicate that the
grains towards AZV 456 have average sizes close to those in the
Galactic diffuse ISM, although with a narrower size distribution.
Inspection of the line of sight to AZV~456 using the catalog
of SMC IRAS sources of Schwering \& Israel (1989) shows that
there is an extended IR source towards that object. This line of
sight might then indeed not represent the typical line of sight
through the diffuse ISM in the SMC, as discussed earlier in this section.

     The lines of sight with smaller \lmax would evidence smaller
average grain sizes. They are associated with no bump in their UV
extinction curves, that is, with the extinction law which is
considered typical of  the SMC (Pr\'evot et al. 1984). The
correlation between the FUV curvature and linear part of the
extinction curve, suggested by the data (sec. 4.2), indicate
smaller average grain sizes producing the typical SMC extinction.
The smaller average grain sizes inferred from our polarization
data and model fits strengthen this interpretation.

\subsection{The SMC environment and dust models}
\label{discussion_models}

     As outlined in the Introduction, the SMC is a valuable
laboratory for studying several aspects of the evolution of the
dust content of the ISM in galaxies (Westerlund 1989, 1990). The
UV bump in the Galactic extinction curve is thought to be
produced by carbon (graphite) grains. Consequently, the lack of
the bump in the SMC extinction curve is usually associated with a
small quantity of carbon grains. However, there is no evidence
for a carbon deficiency in the ISM of the SMC. Despite the
comparison between stellar dust injection and depletion by
star formation and shocks in the interstellar medium, many of the
grains containing carbon are thought to come from carbon stars
(e.g., Gehrz 1989). The relative number of carbon stars to normal
stars in the post main-sequence stages of stellar evolution is
actually greater in the SMC than in the Milky Way (Lequeux 1988
and references therein). Also, the C/Si ratio seems to be
comparable to the Galactic one (Dufton et al. 1990, see Sec. 3).
On the other hand, it has been argued that amorphous carbon
grains are a more plausible interstellar component than graphite
(Bussoletti, Colangeli \& Orofino 1988 and references therein).
Studies of interplanetary dust show that graphite is present only
in trace amounts (Nuth 1985). Amorphous carbon has also been
detected around R CrB stars (Hecht et al. 1984).

     The typical SMC extinction curve, shown by AZV398, can be
best fit with amorphous carbon grains rather than with graphite grains
(sec. \ref{av398}). Moreover, with graphite the required Si
abundance is somewhat greater than that available in the SMC.
Support for
an amorphous carbon component to the AZV~398 extinction is
additionally provided
by the star HD~89353 (=HR~4049) whose extinction curve has no
bump but a high, quasi-linear rise into the UV produced by grains
in a circumstellar envelope (Buss, Lamers \& Snow 1989). Among
the stars in the Galaxy it is one of the poorest in metals and
richest in hydrogen and carbon. Blanco, Fonti \& Orofino (1995)
were able to fit its extinction using optical constants of
amorphous carbon rich in H. On the other hand, Mennella et al.
(1995) show that initially H rich amorphous carbon grains
progressively lose their H as annealing occurs. The extinction
bump varies in position and strength according to the annealing 
temperature. How could H-rich amorphous carbon survive in the SMC
ISM, an environment with a high UV ISRF? What is the role of the
ISM metallicity in grain evolution? Jenniskens et al. (1993)
suggested that an icy mantle formed in molecular clouds can be
transformed to an organic residue due to action of the FUV
radiation, and that a prolonged exposure to this radiation can
lead to the formation of amorphous aromatic carbon.

     In the line of sight to AZV456, which seems to have larger
grains, there is an IR emission associated with cold dust. This
can be considered, therefore, as a region with density
enhancement, similar to the Galactic molecular clouds. The
small gas-to-dust ratio towards this region could also be an
indication of the presence of molecular hydrogen (Lequeux 1994).
If this indeed happens, we can expect that the enhanced density
will lead to the accretion of smaller grains on to the larger
ones. This will increase the average size and reduce the width of
the grain size distribution, as seen in the AZV456 results. If we
believe that (part of) the line of sight to AZV 456 is really characterized by
higher densities, then the grains could be shielded from the
strong ISRF, and the carriers of the 2175\AA\ bump could survive. 
Recently, Andersson \& Wannier (1995) found a Galactic molecular
cloud whose dust is characterized by normal values of \lmax and
large K.

     It has been suggested (Sorrel 1990) that carbon is
transformed into graphite in the ISM by an annealing process
caused by the interstellar radiation field (ISRF).  The SMC
would, in that case, be a privileged site for graphite formation
because the ISRF is greater there than in the Galaxy (Lequeux
1989). But that does not seem to be the case, since the UV bump
is not generally observed.

     Actually, the higher ISRF in the SMC might have a bearing on
the absence of a UV bump in the SMC typical extinction curve.
Leene \& Cox (1987) found a correlation between the
60\mic/100\mmic brightness ratio, a measure of dust temperature,
and the height of the Galactic UV bump in the sense that the bump
gets weaker when the 60\mic/100\mmic ratio (and presumably the
ISRF) gets higher. They suggested that the particles
responsible for the UV bump are very sensitive to the ISRF
intensity. Jenniskens \& Greenberg (1993) also found that the
bump is sensitive to strong UV radiation fields from their
studies of Galactic extinction curves and the environment.

     We now wish to point out some implications of the SMC data
and the hypothesis of Polycyclic Aromatic Hydrocarbons (PAHs).
Jenniskens, Ehrefreund \& D\'esert (1992) have found a
correlation between the amount of FUV rise ($c_4$, sec. 4.1) and
the CH abundance in the line of sight to several Galactic stars.
Since CH is directly proportional to molecular hydrogen, they
suggested that the carrier of the FUV rise coexists with the 
medium containing H$_2$. They also suggested that the FUV rise
is associated with the PAHs. As we have seen, AZV~456 is related
to an IR extended source and its
smaller gas-to-dust ratio suggests the presence of H$_2$.
However, it is precisely AZV~456 which shows a small value
of $c_4$ (Table \ref{tab_fm} and Fig. \ref{fig_cof}).

     A related question concerns the PAH hypothesis and the IR
emission of the SMC. Sauvage, Thuan \& Vigroux (1990) presented
the colors of the IRAS IR emission of the LMC and SMC and
correlated them with the age and metallicity of the underlying
stellar populations. Lequeux (1989) has nicely reviewed these and
other IR data and their implications for the interstellar dust in
the Magellanic Clouds. The integrated 12\mmic emission of the SMC
(Rice et al. 1988, Schwering 1988) is specially low among
irregular galaxies, suggesting that the PAHs, believed to be
responsible for such emission, are less abundant, perhaps having
been destroyed by photodissociation by the more abundant FUV
photons. This could be due to the larger ISRF in the SMC. The
IRAS color-color diagrams suggest that the SMC may be rich in
small (\ltsim 0.05\mic) grains responsible for the extinction
shortwards of 2000\AA. While the SMC extinction and the
polarization data we provide herein support the abundance of
small grains, the steep FUV extinction and low 12\mmic emission
seem to rule out the PAHs as suggested by D\'esert, Boulanger \&
Puget (1990).

\section{Summary}

     In order to study the grains in the SMC, which has an
extinction curve very different from that for the Galaxy, we have
obtained the first wavelength dependent polarization measurements
for stars in the SMC. From an analysis of these and the
extinction curves, for which we present some new IUE data, for a
small sample of stars we reach the following conclusions:

\begin{itemize}

\item the wavelength of maximum polarization, \llmax, determined from a fit
of the Serkowski curve to the new polarization data, is generally
smaller than in the Galaxy;

\item for AZV 456, the single
well-studied case which shows the extinction bump at 2175 \AA,
\lmax is typical of that in the Galaxy; at the same time the bump
for this star is shifted to the blue, as compared to the Galaxy,
and its polarization curve is narrower;

\item  the FUV curvature and
the linear component of the extinction curve increase
simultaneously for stars with small \llmax.

\end{itemize}

We attempt to fit both the extinction and polarization data
for stars in the SMC by varying the mean size and the size
distribution of the grains within the framework of the MRN model
and introducing the notion of volume continuity. Our results are:

\begin{itemize}

\item The AZV~398 (i.e., the typical SMC) extinction curve
is best fit using, in addition to
silicates, amorphous carbon instead of graphite. The polarization data
and the polarization and extinction fits
indicate smaller grains in the SMC as compared to the Galaxy.
Contrary to earlier belief, silicate alone cannot provide
the amount of extinction observed towards the SMC,
despite its providing an approximately correct wavelength dependence;

\item It appears that for AZV456 there may be a contribution, not yet
accounted for, of H$_2$ to the actual ratio, N(H)/A(V).

\end{itemize}

Our results indicate that the typical
line-of-sight in the SMC is characterized by grains smaller and
of a different type on the average than
those in the Galaxy. This may be further evidence that the dust
grains have formed and evolved somewhat differently and/or at a different
rate in the SMC environment than they have in the Galaxy.

\acknowledgments

This work has been supported by NASA Grant NAG 5 1463. It has
been  also partially supported by FAPESP (CVR: 89/3091-6; AMM:
89/1670-9, 92/3345-0 and 94/0033-3) and CNPq (AMM: 301558/79-5).
AMM and CVR would like to acknowledge the hospitality provided by
Dr. A. Code, Space Astronomy Laboratory and Astronomy Department,
University of Wisconsin, where this research was partly done. We
would like to thank the referee, Dr. Peter Martin, for his
valuable criticism which helped improve the paper.

%
%
% Tabelas
%

%\begin{table}
%\tablenum{1}
%\label{tab_pol}
%\end{table}
\include{tab1}

%\begin{table}
%\tablenum{2}
%\label{tab_cor}
%\end{table}
\include{tab2}

%\begin{table}
%\tablenum{3a}
%\label{tab_prog}
%\end{table}

%\begin{table}
%\tablenum{3b}
%\label{tab_comp}
%\end{table}
\include{tab3}

%\begin{table}
%\tablenum{4}
%\label{tab_serk}
%\end{table}
\include{tab4}
%\clearpage

\begin{table}[h]
\tablenum{5}
\label{tab_fm}
\hskip -4cm $ $

\vskip -3cm 
\vspace*{-2cm}
\hspace*{-2.5cm}
\epsfig{file=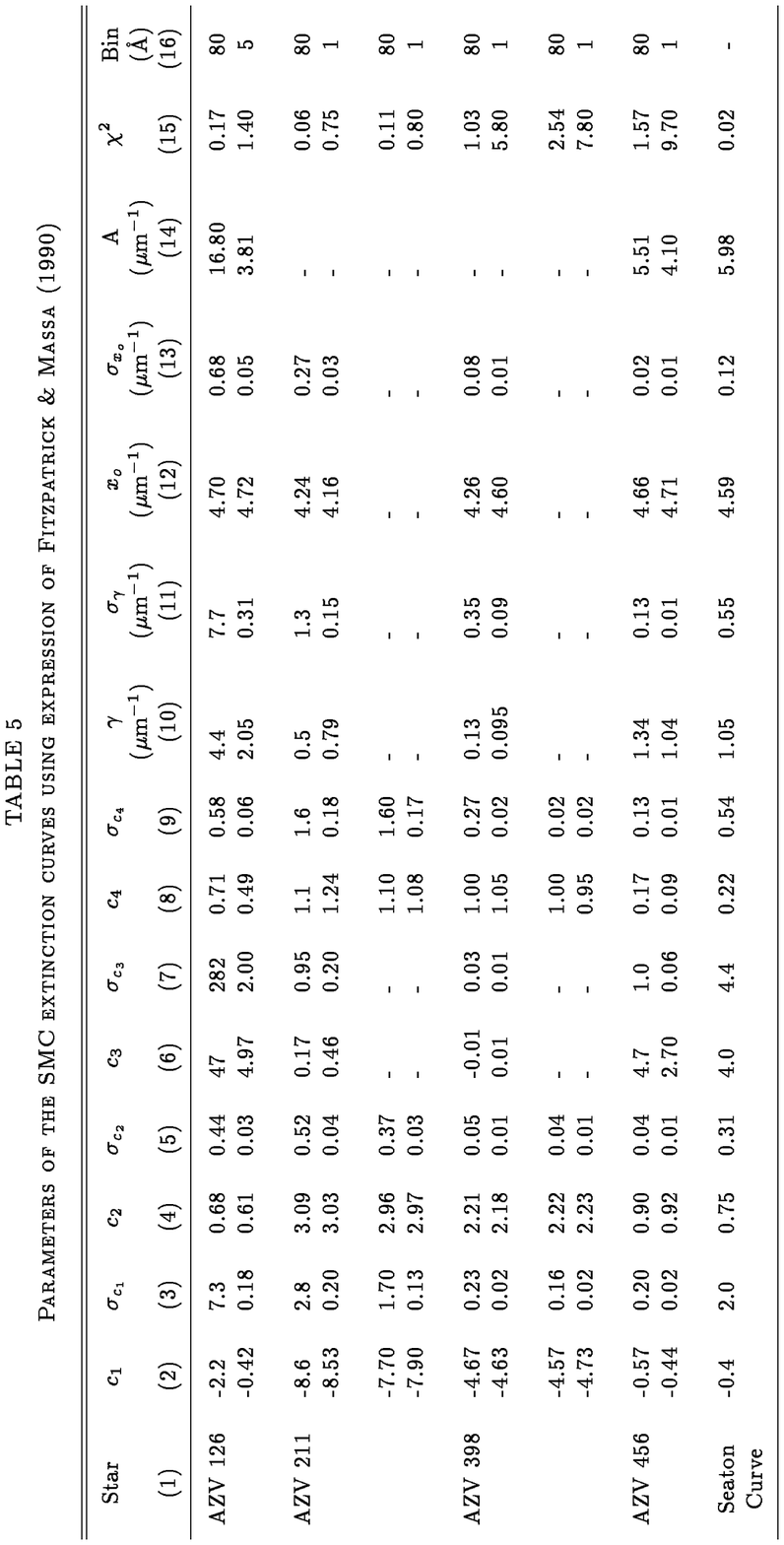,angle=180}
\end{table}
\clearpage

%\begin{table}[h]
%\tablenum{6}
%\label{aj_pol}
%\end{table}
\include{tab6}

%\clearpage

\begin{table}[h]
\tablenum{7}
\label{aj_ext}
\vspace*{-3cm}
\hspace*{-2.5cm}
\epsfig{file=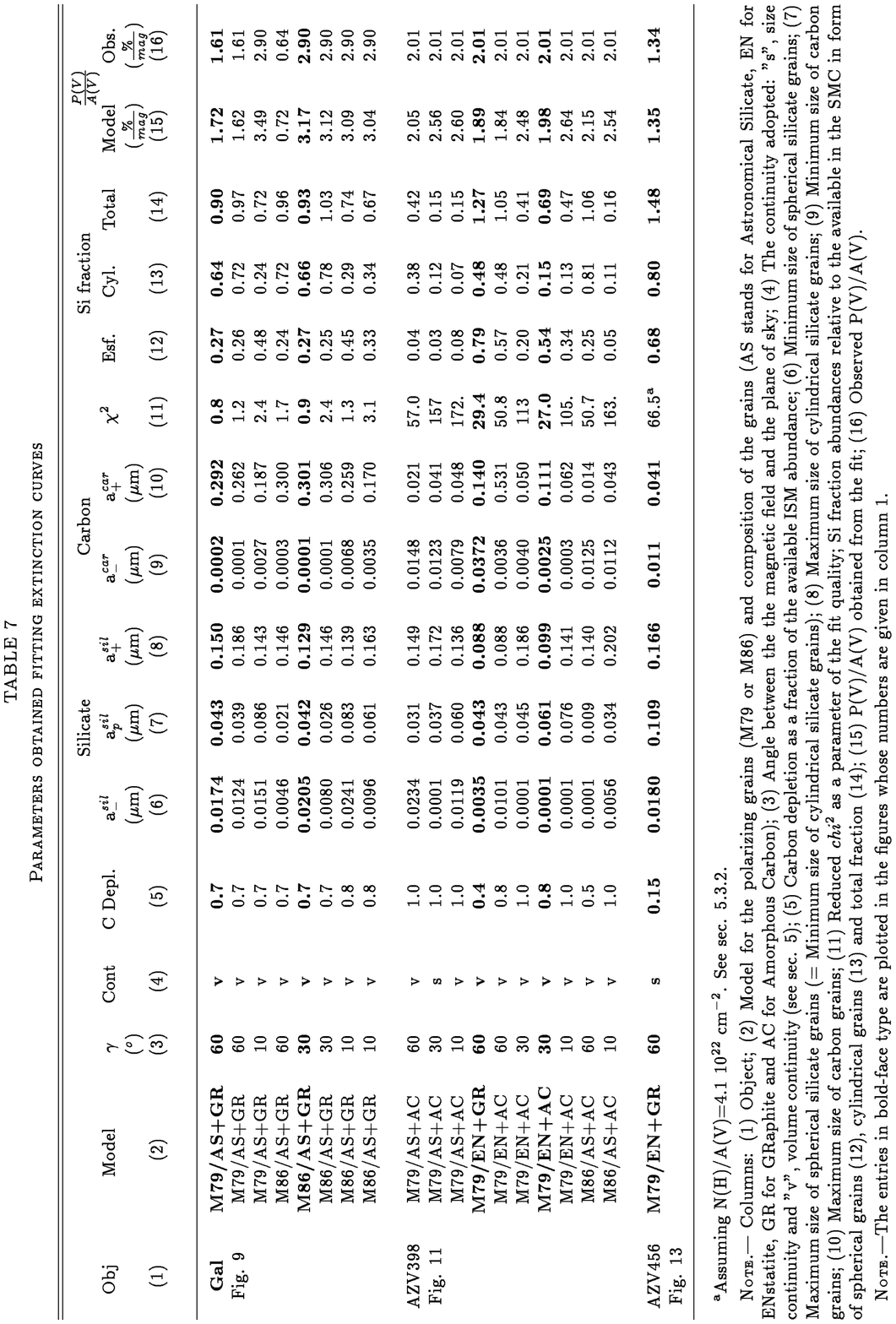,angle=180}
\end{table}
\clearpage

%
%
% Now comes the reference list.  In this document, we used \cite to call
% out citations, so we must use \bibitem in the reference list, which
% means we use the LaTeX thebibliography environment.  Please note that
% \begin{thebibliography} is followed by a null argument.  If you forget
% this, mayhem ensues, and LaTeX will say "Perhaps a missing item?" when
% you run it.  Do not call us, do not send mail when this happens.  Put
% the silly {} after the \begin{thebibliography}.
%
% Each reference has a \bibitem command to define the citation format
% and the symbolic tag, as well as a command which sets up
% formatting parameters for the reference list itself.
%
% If we had not bothered with the \cite-\bibitem business, calling out
% the references outselves, the reference list could be enclosed in
% a references environment (\begin{references} has no null argument),
% and there would be no need for the leading \bibitem's.

%\setcounter{page}{35}

%
%
% Finally, we have figure captions.  Usually these must be on a separate
% page, although unlike table, it is often permissible to have several
% figure captions on the same page.  We force the page break between
% the reference list and the figure captions.
%
% The \caption command in the figure environment works like the one in the
% table environment (it's the same one, actually), except that this one
% produces identification text that reads "Figure N."

\clearpage

\clearpage

% That's all, folks.
%
% The technique of segregating major semantic components of the document
% within "environments" is a very good one, but you as an author have to
% come up with a way of making sure each \begin{whatzit} has a corresponding
% \end{whatzit}.  If you miss one, LaTeX will probably complain a great
% deal during the composition of the document.  Occasionally, you get away
% with it right up to the \end{document}, in which case, you will see
% "\begin{whatzit} ended by \end{document}".

\end{document}

%% file: tab1.tex
% SAMPLE2.TEX -- AASTeX macro package tutorial paper.

% The first item in a LaTeX file must be a \documentstyle command to
% declare the overall style of the paper.  The two \documentstyle lines
% that are relevant for the AASTeX macros are shown; one is commented out
% so that the file can be processed.
% ************ Comentar out esta linha quando compilando com o artigo *****
%\documentstyle[12pt,aasms4]{article}
%\documentstyle[12pt,apjpt4]{article}
%\documentstyle[12pt]{article}
%\usepackage{apjpt}
%
% ************ Comentar out esta linha quando compilando com o artigo *****
%\newcommand{\lmax}{$\lambda_{max}$}
%\newcommand{\mic}{$\micron$}
%
%\tighten
% ************ Comentar out esta linha quando compilando com o artigo *****
%\begin{document}
%
% Tabela: Polarizacoes observada e corrigida do foreground
%
\begin{planotable}{ccccccccc}
\tablenum{1}
\tablecaption{SMC polarization data}
\tablehead{
\multicolumn{2}{c}{Identification} &
\multicolumn{3}{c}{Observed} &
\multicolumn{3}{c}{Corrected} & \nl
\colhead{AZV}  &
\colhead{SK}   &
\colhead{P}     &
\colhead{$\sigma_P$} &
\colhead{$\theta_P$} &
\colhead{P}     &
\colhead{$\sigma_P$} &
\colhead{$\theta_P$} &
\colhead{$\lambda_{eff}$} \nl
& & (\%) & (\%) & ($^o$) & (\%) & (\%) & ($^o$) & (\mic) \nl
\colhead{(1)}		& \colhead{(2)}		&
\colhead{(3)}		& \colhead{(4)}		&
\colhead{(5)}		& \colhead{(6)}		&
\colhead{(7)}		& \colhead{(8)}		&
\colhead{(9)}
}
\startdata 
126 & - & 1.149 & 0.095 & 107.5 & 0.737 & 0.126 & 104.1 & 0.379 \nl
& & 1.175 & 0.055 & 109.3 & 0.725 & 0.105 & 106.6 & 0.439 \nl
& & 1.076 & 0.040 & 111.4 & 0.598 & 0.102 & 109.7 & 0.559 \nl
& & 0.781 & 0.078 & 110.1 & 0.325 & 0.129 & 105.3 & 0.664 \nl
& & 0.825 & 0.083 & 108.6 & 0.407 & 0.120 & 103.7 & 0.791 \nl
\nl
211 & 74 & 1.207 & 0.044 & 126.6 & 0.943 & 0.085 & 127.3 & 0.378 \nl
& & 1.243 & 0.027 & 124.8 & 0.955 & 0.083 & 125.0 & 0.438 \nl
& & 1.250 & 0.075 & 126.0 & 0.947 & 0.112 & 126.6 & 0.516 \nl
& & 1.175 & 0.017 & 126.1 & 0.873 & 0.085 & 126.8 & 0.559 \nl
& & 1.063 & 0.018 & 127.6 & 0.774 & 0.082 & 128.9 & 0.664 \nl
& & 0.986 & 0.042 & 131.1 & 0.728 & 0.085 & 133.7 & 0.790 \nl
& & 0.991 & 0.080 & 125.0 & 0.726 & 0.107 & 125.3 & 0.820 \nl
\nl
215 & 76 & 0.782 & 0.035 & 148.5 & 0.626 & 0.086 & 158.3 & 0.437 \nl
& & 0.907 & 0.032 & 148.6 & 0.741 & 0.089 & 157.4 & 0.557 \nl
& & 0.742 & 0.033 & 148.5 & 0.587 & 0.087 & 159.2 & 0.662 \nl
& & 0.653 & 0.060 & 148.7 & 0.512 & 0.095 & 160.0 & 0.791 \nl
\nl
221 & 77 & 1.125 & 0.099 & 138.3 & 0.903 & 0.129 & 148.4 & 0.377 \nl
& & 1.048 & 0.047 & 140.9 & 0.862 & 0.101 & 153.3 & 0.437 \nl
& & 1.014 & 0.042 & 141.5 & 0.838 & 0.103 & 155.2 & 0.558 \nl
& & 0.799 & 0.060 & 144.3 & 0.698 & 0.109 & 161.6 & 0.663 \nl
& & 0.659 & 0.084 & 143.0 & 0.557 & 0.119 & 162.5 & 0.791 \nl
\nl
398 & - & 2.114 & 0.108 & 131.4 & 1.856 & 0.130 & 132.5 & 0.380 \nl
& & 2.084 & 0.058 & 132.5 & 1.810 & 0.098 & 133.8 & 0.440 \nl
& & 2.038 & 0.042 & 131.9 & 1.751 & 0.093 & 133.2 & 0.560 \nl
& & 1.948 & 0.058 & 134.6 & 1.679 & 0.099 & 136.4 & 0.665 \nl
& & 1.464 & 0.072 & 134.8 & 1.218 & 0.103 & 137.0 & 0.791 \nl
\nl
456 & 143 & 0.959 & 0.074 & 163.9 & 0.842 & 0.086 & 167.0 & 0.376 \nl
& & 1.147 & 0.060 & 160.0 & 1.006 & 0.076 & 162.3 & 0.437 \nl
& & 1.323 & 0.172 & 156.7 & 1.156 & 0.179 & 158.4 & 0.516 \nl
& & 1.338 & 0.045 & 160.9 & 1.195 & 0.067 & 163.0 & 0.559 \nl
& & 1.259 & 0.050 & 160.7 & 1.120 & 0.069 & 162.9 & 0.671 \nl
& & 1.133 & 0.059 & 161.3 & 1.006 & 0.073 & 163.6 & 0.791 \nl
& & 0.950 & 0.089 & 163.2 & 0.830 & 0.099 & 166.2 & 0.820 \nl
\enddata
\label{tab_pol}
\end{planotable}
% ************ Comentar out esta linha quando compilando com o artigo *****
%\end{document}

%% file: tab2.tex
% SAMPLE2.TEX -- AASTeX macro package tutorial paper.
% The first item in a LaTeX file must be a \documentstyle command to
% declare the overall style of the paper.  The two \documentstyle lines
% that are relevant for the AASTeX macros are shown; one is commented out
% so that the file can be processed.
% ************ Comentar out esta linha quando compilando com o artigo *****
%\documentstyle[12pt,aasms]{article}
%\documentstyle[12pt,apjpt4]{article}
%\documentstyle[12pt]{article}
%\usepackage{apjpt}
%\newcommand{\lmax}{$\lambda_{max}$}
%\newcommand{\mic}{$\micron$}
%\tighten
% ************ Comentar out esta linha quando compilando com o artigo *****
%\begin{document}
% Tabela: Polarizacao de foreground
%
\begin{planotable}{cllllc}
\tablenum{2}
\tablecaption{Foreground polarization corrections for various
regions in the SMC}
\tablehead{
\colhead{Region}  &
\colhead{P}   &
\colhead{$\sigma_P$}  &
\colhead{$\theta$} &
\colhead{$\sigma_\theta$} &
\colhead{N} \nl
& (\%) & (\%) & (deg) & (deg) &   \nl
}
\startdata 
I & 0.47 & 0.09 & 113.6 & 14.0 & 6 \nl
\nl
II & 0.30 & 0.08 & 124.1 & 8.0 & 10 \nl
\nl
III & 0.17 & 0.05 & 145.0 & 8.2 & 11 \nl
\nl
IV & 0.22 & 0.05 & 124.2 & 6.5 & 6 \nl
\nl
V & 0.27 & 0.05 & 95.6 & 5.5 & 11 \nl
\nl
\enddata
\label{tab_cor}
\end{planotable}

% ************ Comentar out esta linha quando compilando com o artigo *****
%\end{document}

%% file: tab3.tex
% SAMPLE2.TEX -- AASTeX macro package tutorial paper.
%
%****************************************************************
% ********    Esta tabela chama-se TAB_IUE1.tex    **************
%****************************************************************
%
% The first item in a LaTeX file must be a \documentstyle command to
% declare the overall style of the paper.  The two \documentstyle lines
% that are relevant for the AASTeX macros are shown; one is commented out
% so that the file can be processed.
%
% ************ Comentar out esta linha quando compilando com o artigo *****
%\documentstyle[apjpt4]{article}
% ************ Comentar out esta linha quando compilando com o artigo *****
%\begin{document}
%
% ************ Comentar out estas linhas quando compilando com o artigo *****
%\newcommand{\lmax}{$\lambda_{max}$}
%\newcommand{\mic}{$\micron$}
%
%\tighten
%
%***** Tabela com dados das estrelas avermelhadas *****
%
\begin{planotable}{llllllll}
\tablenum{3a}
\tablecaption{Relevant data on the reddened stars} 
\tablehead{
\colhead{Star}  &
\colhead{Sp}  &
\colhead{$(B-V)_0$\tablenotemark{f}} &
\colhead{V}\tablenotemark{g} &
\colhead{(B-V)} &
\colhead{(U-B)} &
\colhead{E(B-V)} &
\colhead{IUE Images} 
}
\startdata 
AZV~20 & $A0Ia\tablenotemark{a}$ & -0.04 & 12.1 & +0.29 & -0.12 & +0.33 & LWP
21313,14,15 \nl
& & & & & & & SWP 42509, 39,40 \nl
\nl
AZV~126 & $B0Iw\tablenotemark{b}$ & -0.25 & 13.47 & -0.02 & -0.90 & +0.23 & LWP
21279,80 \nl
& & & & & & & LWR 14947 \nl
& & & & & & & SWP 42506, 18908 \nl
\nl
AZV~211 & $A0Ia\tablenotemark{d}$ & -0.04 & 11.5 & +0.10 & -0.45 & +0.14 & LWP
21279, 80 \nl
& & & & & & & LWR 14947 \nl
& & & & & & & SWP 42506, 18908 \nl
\nl
AZV~398 & $B2\tablenotemark{e}$ & -0.18 & 13.85 & +0.09 & -0.77 & +0.27 & LWR
14963
\nl
& & & & & & +0.37\tablenotemark{h} & SWP 18911 \nl
\nl
AZV~456 & $B0-1\tablenotemark{e}$ & -0.25 & 12.89 & +0.10 & -0.74 & +0.35 & LWR
12347 \nl
& & & & & & +0.36\tablenotemark{h} & SWP 16051 \nl
\nl
\enddata
\label{tab_prog}
\end{planotable}

% ***** Tabela: Detalhes sobre as estrelas de comparacao da SMC *****
%
\begin{planotable}{lllllcll}
\tablenum{3b}
\tablecaption{Relevant data on the comparison stars} 
\tablehead{
\colhead{Comparison}  &
\colhead{Reddened}  &
\colhead{Sp} &
\colhead{$(B-V)_0$\tablenotemark{f}} &
\colhead{V}\tablenotemark{g} &
\colhead{(B-V)} &
\colhead{(U-B)} &
\colhead{IUE Images} \nl
\colhead{Star} & \colhead{Star} & & & & & & 
}
\startdata 
AZV~61 & AZV~126 & $O5V\tablenotemark{c}$ & +0.01 & 13.68 & -0.23 & -0.98 & LWP
19245 \nl
\nl
AZV~161 & AZV~20 & $A0I\tablenotemark{a}$ & +0.01 & 11.80 & +0.03 & -0.40 & LWP
19245 \nl
& AZV~211 & & & & & & SWP 40139 \nl
\nl
AZV~235 & AZV~398 & $B0Iw\tablenotemark{b}$ & -0.24 & 12.15 & -0.12 & -0.94 & LWR
7239, 84 \nl
& AZV~456 & & & & & & SWP 8293 \nl
\nl
AZV~242 & AZV~398 & $B1:I\tablenotemark{a}$ & -0.19 & 12.08 & -0.10 & -0.88 & LWR
7242 \nl
& AZV~456 & & & & & & SWP 8296 \nl
\nl
AZV~270 & AZV~20 & $A0Ia\tablenotemark{d}$ & +0.02 & 11.43 & +0.03 & -0.42 & LWP
19241, 44
\nl
& AZV~211 & & & & & & SWP 40134, 37 \nl
\nl
AZV~289 & AZV~398 & $B0.5I\tablenotemark{c}$ & -0.22 & 12.42 & -0.14 & -0.94 & LWR
12345 \nl
& AZV~456 & & & & & & SWP 16049, 18829 \nl
\nl
AZV~317 & AZV~126 & $B0Iw\tablenotemark{b}$ & -0.24 & 12.90 & -0.20 & -1.00 & LWR
17264 \nl
& AZV~398 & & & & & & SWP 10315, 22373 \nl
& AZV~456 & & & & & & \nl
\nl
AZV~454 & AZV~126 & $OV\tablenotemark{e}$ & - & - & -0.19 & -0.98 & LWR
14948 \nl
& & & & & & & SWP 18909, 22016 \nl
\nl
AZV~488 & AZV~398 & $B0Ia\tablenotemark{d}$ & -0.24 & 11.88 & -0.13 & -0.97 & LWR
5642 \nl
& AZV~456 & & & & & & SWP 6590 \nl
\nl
AZV~504 & AZV~20 & $B9Ia\tablenotemark{d}$ & - & 11.91 & -0.04 & -0.46 & LWP 19242
\nl
& AZV~211 & & & & & & SWP 40138 \nl
\nl
SK~194 & AZV~20 & $B9Ia\tablenotemark{d}$ & - & 11.74 & +0.02 & -0.53 & LWP
21282, 83 \nl
& AZV~211 & & & & & & SWP 42507, 08 \nl
\nl
\tablenotetext{a}{Humphreys (1983)}
\tablenotetext{b}{Garmany et al. (1987)}
\tablenotetext{c}{Crampton \& Greasley (1982)}
\tablenotetext{d}{Ardeberg et al. (1972)}
\tablenotetext{e}{Prevot et al. (1984)}
\tablenotetext{f}{Brunet (1975)}
\tablenotetext{g}{Photometry from AZV82}
\tablenotetext{h}{Bouchet et al. (1985)}
\enddata
\label{tab_comp}
\end{planotable}

% ************ Comentar out esta linha quando compilando com o artigo *****
%\end{document}

%% file: tab4.tex
% SAMPLE2.TEX -- AASTeX macro package tutorial paper.

% The first item in a LaTeX file must be a \documentstyle command to
% declare the overall style of the paper.  The two \documentstyle lines
% that are relevant for the AASTeX macros are shown; one is commented out
% so that the file can be processed.
%
% ************ Comentar out esta linha quando compilando com o artigo *****
%\documentstyle[apjpt4]{article}
%
%\newcommand{\lmax}{$\lambda_{max}$}
%
% ************ Comentar out esta linha quando compilando com o artigo *****
%\newcommand{\mic}{$\micron$}
%
%\tighten
%
% ************ Comentar out esta linha quando compilando com o artigo *****
%\begin{document}
%
% Tabela: Ajustes da pol. - serkowski
%
\begin{planotable}{cllllllllc}
\tablenum{4}
\tablecaption{Parameters of the Serkowski curve from fits of the SMC 
polarization data}
\tablehead{
\colhead{Star}  &
\colhead{$\lambda_{max}$}  &
\colhead{$\sigma_{\lambda_{max}}$} &
\colhead{$K$} &
\colhead{$\sigma_K$} &
\colhead{$P_{max}$}  &
\colhead{$\sigma_{P_{max}}$} &
\colhead{$\chi_{\nu}^2$} &
\colhead{$P(\chi_{\nu}^2)$} &
\colhead{$\nu$}\nl
& (\mic) & (\mic) & & & (\%) & (\%) \nl
}
\startdata 
AZV~126 & 0.31 & 0.31 & 1.0 & 1.8 & 0.79 & 0.34 & 0.73 & 0.48 & 2\nl
& 0.20 & 0.99 & 0.3 & 1.6 & 0.8 & 1.4 & 0.77 & 0.51 & 3\nl
\nl
AZV~211 & 0.37 & 0.22 & 0.48 & 0.67 & 0.956 & 0.089 & 0.11 & 0.98 & 4\nl
& 0.42 & 0.06 & 0.70 & 0.10 & 0.950 & 0.053 & 0.11 & 0.99 & 5\nl
\nl
AZV~215 & 0.54 & 0.05 & 2.5 & 1.9 & 0.707 & 0.071 & 0.54 & 0.46 & 1\nl
& 0.48 & 0.12 & 0.81 & 0.20 & 0.666 & 0.061 & 0.66 & 0.52 & 2 \nl
\nl
AZV~221 & 0.42 & 0.12 & 1.2 & 1.3 & 0.892 & 0.073 & 0.09 & 0.91 & 2\nl
& 0.34 & 0.14 & 0.57 & 0.23 & 0.910 & 0.140 & 0.13 & 0.94 & 3\nl
\nl
AZV~398 & 0.46 & 0.04 & 1.26 & 0.55 & 1.87 & 0.07 & 1.15 & 0.32 & 2\nl
& 0.40 & 0.05 & 0.68 & 0.08 & 1.86 & 0.08 & 1.17 & 0.32 & 3\nl
\nl
AZV~456 & 0.57 & 0.02 & 2.06 & 0.53 & 1.19 & 0.05 & 0.30 & 0.88 & 4\nl
& 0.59 & 0.03 & 0.98 & 0.06 & 1.11 & 0.03 & 1.10 & 0.36 & 5\nl
\enddata
\label{tab_serk}
\end{planotable}

% ************ Comentar out esta linha quando compilando com o artigo *****
%\end{document}

%% file: tab6.tex
% SAMPLE2.TEX -- AASTeX macro package tutorial paper.
% The first item in a LaTeX file must be a \documentstyle command to
% declare the overall style of the paper.  The two \documentstyle lines
% that are relevant for the AASTeX macros are shown; one is commented out
% so that the file can be processed.
%\input{psfig.sty}
%\documentstyle[12pt,apjpt4]{article}
%\documentstyle[11pt,aaspp]{article}
%\documentstyle[aaspptwo]{article}
%
%\newcommand{\lmax}{$\lambda_{max}$}
%\newcommand{\mic}{$\micron$}
%
%\tighten
%\clearpage
%\setcounter{page}{32}
%\begin{document}
%
%Table 2: Parameters obtained from fits to the polarimetric data for the SMC
% and the Galaxy  
%
\begin{deluxetable}{cccccccccc}
%\tablewidth{6.5in}
\tablenum{6}
\tablecaption{Parameters obtained from fits to the polarimetric data for the SMC
and the Galaxy}
\tablehead{
\colhead{Object/} & \colhead{a$_{p}^{sil}$} & \colhead{a$_{+}^{sil}$} &
\colhead{$\langle a\rangle$} & \colhead{Width} & 
\colhead{$\delta (Si)$\tablenotemark{a}} & \colhead{Model} & \colhead{Mat.} &
\colhead{$\gamma$} & \colhead{$\chi^2$} \nl
Figure & \colhead{(\mic )} & \colhead{(\mic )} & \colhead{(\mic )} &
\colhead{(\mic)} & \colhead{(dex)} & & & \colhead{($^o$)} & \nl
\colhead{(1)} & \colhead{(2)} &
\colhead{(3)} & \colhead{(4)} &
\colhead{(5)} & \colhead{(6)} &
\colhead{(7)} & \colhead{(8)} &
\colhead{(9)} & \colhead{(10)}}
\startdata
Gal & 0.043 & 0.150 & 0.063 & 0.107 & 7.61 & M79 & AS & 60 & 0.00003 \nl
Fig. 8 & 0.039 & 0.186 & 0.059 & 0.147 & 7.66 & M79 & AS & 60 & 0.00293 \nl
& {\bf 0.054} & {\bf 0.162} & {\bf 0.078} & {\bf 0.107} & {\bf 7.09} & {\bf M79} 
& {\bf AS} & {\bf 30} & {\bf 0.00035} \nl
& {\bf 0.050} & {\bf 0.190} & {\bf 0.074} & {\bf 0.140} & {\bf 7.13} & {\bf M79} 
& {\bf AS} & {\bf 30} & {\bf 0.00041} \nl
& {\bf 0.045} & {\bf 0.241} & {\bf 0.069} & {\bf 0.197} & {\bf 7.19} & {\bf M79} 
& {\bf AS} & {\bf 30} & {\bf 0.00267} \nl
& 0.062 & 0.203 & 0.090 & 0.141 & 6.98 & M79 & AS & 10 & 0.00023 \nl
& 0.062 & 0.198 & 0.090 & 0.136 & 6.98 & M79 & AS & 10 & 0.00027 \nl
& 0.086 & 0.143 & 0.106 & 0.057 & 6.93 & M79 & AS & 10 & 0.00356 \nl
& 0.021 & 0.146 & 0.062 & 0.125 & 8.06 & M86 & AS & 60 & 0.02490 \nl
& 0.015 & 0.146 & 0.057 & 0.131 & 8.10 & M86 & AS & 60 & 0.02406 \nl
& 0.010 & 0.146 & 0.052 & 0.136 & 8.14 & M86 & AS & 60 & 0.02394 \nl
& 0.006 & 0.146 & 0.048 & 0.140 & 8.17 & M86 & AS & 60 & 0.02395 \nl
& 0.042 & 0.129 & 0.075 & 0.087 & 7.37 & M86 & AS & 30 & 0.01162 \nl
& 0.026 & 0.146 & 0.066 & 0.120 & 7.44 & M86 & AS & 30 & 0.01045 \nl
& 0.020 & 0.148 & 0.061 & 0.128 & 7.48 & M86 & AS & 30 & 0.01037 \nl
& 0.010 & 0.148 & 0.052 & 0.138 & 7.54 & M86 & AS & 30 & 0.01031 \nl
& 0.083 & 0.139 & 0.106 & 0.056 & 7.01 & M86 & AS & 10 & 0.00312 \nl
& 0.061 & 0.163 & 0.095 & 0.101 & 7.08 & M86 & AS & 10 & 0.00058 \nl
& 0.030 & 0.179 & 0.073 & 0.149 & 7.22 & M86 & AS & 10 & 0.00076 \nl
\nl
AZV398 & 0.031 & 0.149 & 0.048 & 0.119 & 6.46 & M79 & AS & 60 & 1.62 \nl
Fig. 10 & {\bf 0.037} & {\bf 0.172} & {\bf 0.057} & {\bf 0.135} & {\bf 5.96} &
{\bf  M79} & {\bf AS} & {\bf 30} & {\bf 1.57} \nl
 & 0.060 & 0.136 & 0.081 & 0.076 & 5.75 & M79 & AS & 10 & 1.11 \nl
 & 0.043 & 0.088 & 0.056 & 0.046 & 6.56 & M79 & ENS & 60 & 0.34 \nl
 & 0.045 & 0.186 & 0.069 & 0.141 & 6.21 & M79 & ENS & 30 & 1.50 \nl
 & {\bf 0.061} & {\bf 0.099} & {\bf 0.075} & {\bf 0.038} & {\bf 6.07} & 
{\bf M79} & {\bf ENS} & {\bf 30} & {\bf 0.34} \nl
 & 0.076 & 0.141 & 0.098 & 0.065 & 5.99 & M79 & ENS & 10 & 1.26 \nl
 & 0.031 & 0.082 & 0.047 & 0.051 & 6.64 & M86 & AS & 60 & 0.48 \nl
 & 0.009 & 0.140 & 0.032 & 0.131 & 6.79 & M86 & AS & 60 & 2.36 \nl
 & 0.031 & 0.168 & 0.053 & 0.137 & 6.10 & M86 & AS & 30 & 1.63 \nl
 & 0.034 & 0.202 & 0.058 & 0.168 & 5.94 & M86 & AS & 10 & 1.69 \nl
 & 0.031 & 0.162 & 0.054 & 0.131 & 6.89 & M86 & ENS & 60 & 1.71 \nl
 & 0.042 & 0.089 & 0.057 & 0.047 & 6.77 & M86 & ENS & 60 & 0.59 \nl
 & 0.042 & 0.195 & 0.066 & 0.153 & 6.31 & M86 & ENS & 30 & 1.58 \nl
 & {\bf 0.062} & {\bf 0.098} & {\bf 0.075} & {\bf 0.036} & {\bf 6.14} & {\bf 
M86} & {\bf ENS} & {\bf 30} & {\bf 0.38} \nl
 & 0.078 & 0.136 & 0.098 & 0.058 & 5.99 & M86 & ENS & 10 & 1.44 \nl
 & \nl 
AZV456 & 0.045 & 0.145 & 0.066 & 0.108 & 7.10 & M79 & AS & 60 & 6.43 \nl
Fig. 12 & 0.056 & 0.160 & 0.080 & 0.104 & 6.58 & M79 & AS & 30 & 7.24 \nl
 & {\bf 0.099} & {\bf 0.131} & {\bf 0.112} & {\bf 0.032} & {\bf 6.40} & 
{\bf M79} & {\bf AS} & {\bf 10} & {\bf 1.93} \nl
 & {\bf 0.109 } & {\bf 0.166 } & {\bf 0.131 } & {\bf 0.057 } & {\bf 7.56 } & 
{\bf M79 } & {\bf ENS } & {\bf 60 } & {\bf 2.21} \nl
 & 0.047 & 0.202 & 0.072 & 0.155 & 7.41 & M79 & ENS & 60 & 10.07 \nl
 & 0.131 & 0.173 & 0.149 & 0.042 & 6.99 & M79 & ENS & 30 & 12.94 \nl
 & 0.062 & 0.192 & 0.090 & 0.130 & 6.87 & M79 & ENS & 30 & 10.29 \nl
 & 0.084 & 0.200 & 0.116 & 0.116 & 6.71 & M79 & ENS & 10 & 4.67 \nl
 & {\bf 0.112} & {\bf 0.145} & {\bf 0.126} & {\bf 0.032} & {\bf 6.65} & {\bf 
M79} & {\bf ENS} & {\bf 10} & {\bf 1.72} \nl
 & 0.038 & 0.144 & 0.075 & 0.107 & 7.51 & M86 & AS & 60 & 11.44 \nl
 & 0.047 & 0.157 & 0.084 & 0.111 & 6.87 & M86 & AS & 30 & 10.72 \nl
 & 0.055 & 0.184 & 0.094 & 0.129 & 6.65 & M86 & AS & 10 & 6.52 \nl
\nl
\enddata
 \tablenotetext{a}{The observed values of the Si/H abundance
for the Galaxy (Anders \& Grevesse 1989) and the SMC (Dufton et al. 1990)
are 7.55 and 6.88, respectively.}
\tablecomments{Columns: (1) The object name in the Azzopardi \& Vigneau
(1982) catalogue; (2) The
minimum size of cylindrical grain; (3) The maximum size of
cylindrical grain; 
(4) Average size;
(5) Width of the size distribution which corresponds to the difference
between the two size parameters;
(6) Silicon abundance needed to reproduce the degree of polarization;
(7) Models: M79 (Mathis 1979) or M86 (Mathis 1986);
(8) Material used in the calculations: AS (astronomical silicate) or ENS
(Enstatite). See Sec. 2 for references;
(9) Angle between the magnetic field and the
plane of sky and
(10) The value of $\chi^2$ is not divided by N.}
\tablecomments{The fits in bold-face type are plotted in the figures
whose numbers are given in column 1.}
\label{aj_pol}
\end{deluxetable}

%\end{document}